\documentclass[reprint,aps,pra,superscriptaddress,nofootinbib,longbibliography]{revtex4-1}
\usepackage[latin9]{inputenc}
\pdfoutput=1
\usepackage{color}
\usepackage{times}
\usepackage{amsmath}
\usepackage{amssymb}
\usepackage{stmaryrd}
\usepackage{makecell}
\usepackage[caption=false]{subfig}
\usepackage{graphicx}
\usepackage{epstopdf}
\usepackage{comment}
\usepackage[T1]{fontenc} 
\usepackage[unicode=true,
 bookmarks=true,bookmarksnumbered=false,bookmarksopen=false,
 breaklinks=false,pdfborder={0 0 1},backref=false,colorlinks=true]
 {hyperref}
\hypersetup{
 linkcolor=magenta, urlcolor=blue, citecolor=blue, pdfstartview={FitH}, hyperfootnotes=true, unicode=true}
\makeatletter
\@ifundefined{textcolor}{}
{%
 \definecolor{BLACK}{gray}{0}
 \definecolor{WHITE}{gray}{1}
 \definecolor{RED}{rgb}{1,0,0}
 \definecolor{GREEN}{rgb}{0,1,0}
 \definecolor{BLUE}{rgb}{0,0,1}
 \definecolor{CYAN}{cmyk}{1,0,0,0}
 \definecolor{MAGENTA}{cmyk}{0,1,0,0}
 \definecolor{YELLOW}{cmyk}{0,0,1,0}
}

\renewcommand{\b}[1]{{\mathbf{#1}}}

\newcommand{\sgn}{\mathop{\mathrm{sgn}}}

\newcommand{\etal}{\textit{et al}. }

\newcommand{\bmx}{\begin{pmatrix}}
\newcommand{\emx}{\end{pmatrix}}
\newcommand{\bsmx}{\begin{smallmatrix}}
\newcommand{\esmx}{\end{smallmatrix}}
\newcommand{\vect}[2]{\begin{pmatrix} {#1} \\ {#2} \end{pmatrix}}

\newcommand{\doublewidetilde}[1]{{%
  \mathpalette\double@widetilde{#1}%
}}
\newcommand{\double@widetilde}[2]{%
  \sbox\z@{$\m@th#1\widetilde{#2}$}%
  \ht\z@=.9\ht\z@
  \widetilde{\box\z@}%
}

\usepackage{amsfonts}\usepackage{tabularx}\usepackage{dcolumn}\usepackage{bm}\usepackage{graphicx}\usepackage{epstopdf}

\hypersetup{urlcolor=blue}

\usepackage{multirow}
\usepackage{tensor}
\usepackage{braket}

\usepackage[capitalise,compress]{cleveref}
\crefname{section}{Sec.}{Secs.}
\Crefname{section}{Section}{Sections}
\crefrangelabelformat{equation}{\textup{(#3#1#4)}--\textup{(#5#2#6)}}

\makeatother

\definecolor{darkgreen}{rgb}{0.0, 0.6, 0.13}

\usepackage{amsfonts}\usepackage{tabularx}\usepackage{dcolumn}\usepackage{bm}\usepackage{graphicx}\usepackage{epstopdf}

\hypersetup{urlcolor=blue}

\newcolumntype{x}[1]{%
>{\raggedleft\hspace{0pt}}p{#1}}%

\makeatother
\begin{document}

\title{On the nature of the non-equilibrium phase transition in the non-Markovian driven Dicke model}

\author{Rex Lundgren}
\affiliation{Joint Quantum Institute, NIST/University of Maryland, College Park, MD 20742, USA}

\author{Alexey V. Gorshkov}
\affiliation{Joint Quantum Institute, NIST/University of Maryland, College Park, MD 20742, USA}
\affiliation{Joint Center for Quantum Information and Computer Science, NIST/University of Maryland, College Park, MD 20742, USA}

\author{Mohammad F. Maghrebi}
\affiliation{Department of Physics and Astronomy, Michigan State University, East Lansing, Michigan 48824, USA}

\begin{abstract}
The Dicke model famously exhibits a phase transition to a superradiant phase with a macroscopic population of photons and is realized in multiple settings in open quantum systems.
In this work, we study a variant of the Dicke model where the cavity mode is lossy due to the coupling to a Markovian environment while the atomic mode is coupled to a colored bath. We analytically investigate this model by inspecting its low-frequency behavior via the Schwinger-Keldysh field theory and carefully examine the nature of the corresponding superradiant phase transition. Integrating out the fast modes, we can identify a simple effective theory allowing us to derive analytical expressions for various critical exponents, including those---such as the dynamical critical exponent---that have not been previously considered. We find excellent agreement with previous numerical results when the non-Markovian bath is at zero temperature; however, contrary to these studies, our low-frequency approach reveals that the same exponents govern the critical behavior when the colored bath is at finite temperature unless the chemical potential is zero. Furthermore, we show that the superradiant phase transition is classical in nature, while it is genuinely non-equilibrium. We derive a fractional Langevin equation and conjecture the associated fractional Fokker-Planck equation that capture the system's long-time memory as well as its non-equilibrium behavior. Finally, we consider finite-size effects at the phase transition and identify the finite-size scaling exponents, unlocking a rich behavior in both statics and dynamics of the photonic and atomic observables.
\end{abstract}


\maketitle

\section{Introduction}
Understanding and classifying the nature of non-equilibrium phase transitions in driven-dissipative systems has been a topic of intense active research \cite{2010NatPh...6..806D,PhysRevA.84.043637,PhysRevB.85.184302,1402-4896-2012-T151-014026,PhysRevLett.110.195301,PhysRevA.87.023831,PhysRevA.87.063622,PhysRevLett.115.043601,PhysRevA.94.063862,PhysRevLett.116.070407,PhysRevB.94.085150,PhysRevA.94.061802,2018arXiv180509828K}. This is due in part to the rapid experimental progress in controlling the interplay of coherent driven dynamics and dissipation in atomic, molecular, and optical systems \cite{balili2007bose,peyronel2012quantum,schindler2013quantum,PhysRevLett.116.113001}. 
Remarkably, it has been shown that $1/f$ noise in certain driven-dissipative systems such as a noisy Josephson junction leads to genuinely quantum non-equilibrium critical behavior~\cite{2010NatPh...6..806D} although nonlinear interactions could mask this behavior at larger scales \cite{PhysRevB.85.184302}. Another instance is proposed in a one-dimensional driven system with diffusive Markovian noise~\cite{PhysRevLett.116.070407} where non-equilibrium quantum critical behavior emerges at intermediate scales before the onset of classical behavior at long distances \cite{PhysRevB.94.085150}. It also appears that
some weakly dissipative driven systems give rise to quantum critical behavior \cite{PhysRevLett.122.110405} (though weak dissipation does not always lead to such behavior \cite{2019arXiv190608278P}). 
These examples are in contrast with the typical situation where drive and dissipation together introduce an effective temperature \cite{PhysRevLett.97.236808}, and render the phase transition classical in nature.

Recently, Nagy and Domokos showed that a seemingly similar behavior can occur in a non-Markovian system \cite{PhysRevLett.115.043601,PhysRevA.94.063862}. They theoretically investigated a variant of the driven-dissipative Dicke model where the photonic cavity mode is coupled to a standard Markovian bath while the atomic mode is coupled to a colored bath. Interestingly, the critical exponents were shown to depend on the spectral density of the colored bath. Upon increasing the temperature (of the colored bath) they found the critical exponents changed, but still depended on the spectral density of the colored bath. The authors claimed that this was evidence of a quantum non-equilibrium phase transition.

In this work, we expand on the results of Nagy and Domokos  \cite{PhysRevLett.115.043601,PhysRevA.94.063862}. We first introduce and review the Hamiltonian of the driven Dicke model. We then combine the coherent dynamics with dissipation due to the coupling to the baths within the Schwinger-Keldysh framework. Integrating out the fast (gapped) modes, we then obtain an effective theory describing the low-frequency behavior of either the photonic or the atomic field in terms of a single scalar field, indicative of the Ising character (signified by the $Z_2$ symmetry) of the superradiant phase transition in the Dicke model. We then examine this effective description to identify the critical behavior at large scales and long times. The low-frequency properties of this and other closely related models, including the effective (equilibrium or not) behavior and the (classical or quantum) nature of the corresponding phase transition, are summarized in Table~\ref{situtionTable}. Notably, we find that, when both Markovian and non-Markovian baths are present, the system cannot be described by an effective thermal behavior; however, the non-equilibrium phase transition becomes classical in nature, in contrast to what was previously claimed \cite{PhysRevLett.115.043601}.

\begin{table*}
\begin{tabular}{|c|c|c|c|c|c|}\hline
\makecell{Markovian \\bath} & \makecell{Non-Markovian \\ bath} & \makecell{Temperature and \\ chemical potential \\ of non-Markovian bath} & Distribution function &  \makecell{Effective equilibrium?} & \makecell{Quantum or classical \\ phase transition?} \tabularnewline
\hline
 On & On &\makecell{$T_b=0,~\mu_b\leq0$  \\   $T_b\ne 0,~\mu_b < 0  $}  &  $A/|\omega|^s\sgn(\omega)$& No & Classical  \tabularnewline
\hline 
On & On & \makecell{$T_b\neq 0,~\mu_b=0$} & $2T_b/\omega$& Yes & Classical \tabularnewline
\hline
On &  Off & \makecell{Not Applicable} &$2T_{\rm eff}/\omega$ & Yes & Classical \tabularnewline
\hline
Off  &  On &\makecell{$T_b=0,~\mu_b\leq0$} & $\sgn(\omega)$ & Yes & Quantum \tabularnewline
\hline
\end{tabular}
\caption{Different scenarios considered in this work where the system is coupled to one or both baths and depending on the temperature and chemical potential of the non-Markovian bath; the latter defines a sub-ohmic bath with the spectral density $\rho(\omega)\sim \omega^s$ at low frequencies ($0<s<1$). The low-frequency properties (of the photonic field) are reported near the phase transition. The distribution function (determined by the ratio of correlation and response functions) indicates whether or not the system is in (effective) equilibrium at low frequencies and distinguishes between zero and finite temperature corresponding to quantum and classical phase transitions, respectively. With both baths present and $T_b=0$ and $\mu_b\le 0$, or, alternatively, $T_b\ne 0$ and $\mu_b< 0$, the phase transition is genuinely non-equilibrium ($A$ is a constant depending on microscopic parameters).
With both baths present but $T_b\neq0$ and $\mu_b=0$, an effective thermal behavior emerges where the effective temperature coincides with the temperature of the non-Markovian bath.
In the absence of the non-Markovian bath, a thermal behavior emerges too with the effective temperature $T_{\rm eff}=(\kappa^2+\Delta^2)/4\Delta$, where $\Delta$ and $\kappa$ are the cavity's detuning and decay rate, respectively. When coupled only to the non-Markovian bath at zero temperature, the system equilibrates to zero temperature and exhibits a dissipative quantum phase transition.}
\label{situtionTable}
\end{table*}

Using our low-frequency description, we analytically calculate various critical exponents, including the photon-flux exponent as well as the dynamical critical exponent at and away from criticality. We compare these exponents to the ones obtained numerically by Nagy and Domokos and find excellent agreement when the colored bath is at zero temperature. However, when the colored bath is at finite temperature and finite chemical potential, we find that the critical exponents do not change from their zero temperature values, which nevertheless disagrees with the numerical results in Ref.~\cite{PhysRevA.94.063862}. We explicitly show that this is because one should consider close enough distances to criticality. At finite temperature and zero chemical potential, we find 
critical behavior and exponents consistent with an effective thermal behavior. Finally, we consider the finite-size effects of the spin (atomic degree of freedom). Specifically, we derive the finite-size scaling exponent characterizing the dependence of the photon number on the system size and also identify the finite-size scaling of the dynamics at criticality due to the emergence of a characteristic time scale that diverges algebraically with the system size. Remarkably, in both cases, the results are strongly dependent on the spectral density of the colored bath. These results are presented in Tables~\ref{CritEXPTABLE} and~\ref{crossovertable}. In summary, the main results of this work are presented in Tables~\ref{situtionTable}, \ref{CritEXPTABLE}, and~\ref{crossovertable}.

Our paper is organized as follows. In Sec.~\ref{sec:model}, we review the model originally introduced by Nagy and Domokos in Ref. \cite{PhysRevLett.115.043601}. In Sec.~\ref{sec:LEFT}, we explicitly derive the low-frequency Keldysh field theory of this model by integrating out the fast modes
and argue that the non-equilibrium phase transition found in Ref. \cite{PhysRevLett.115.043601} is classical and not quantum as previously claimed. We show that the effective dynamics is stochastic (due to the coupling to the Markovian bath) and involves fractional derivatives with long-time memory (due to the coupling to the non-Markovian bath). We also conjecture a non-equilibrium fractional Fokker-Planck equation describing the effectively classical dynamics. In Sec.~\ref{sec:CORR}, we use the low-frequency effective theory to analytically calculate various critical exponents and compare them to numerical calculations finding excellent agreement. We also provide numerical evidence that the critical exponents remain identical to their zero temperature value at finite temperature (unless the chemical potential is zero). In Sec.~\ref{sec:othermodels}, we discuss several closely related models and contrast their critical behavior against the main 
model considered in this work. Finally, in Sec.~\ref{sec:FUTURE}, we present a summary of our results and discuss future directions.

\section{Model}\label{sec:model}
In this section, we review the driven-dissipative Dicke model introduced by Nagy and Domokos \cite{PhysRevLett.115.043601}. We first discuss the driven Dicke Hamiltonian and then include the dissipation via the Schwinger-Keldysh action of the system.

\subsection{Driven-Dicke Hamiltonian}

The Hamiltonian for the driven Dicke model used in Refs. \cite{PhysRevLett.115.043601,PhysRevA.94.063862} is given by (in units where Planck's constant, $\hbar$, is unity)
\begin{equation}
H=\omega_0a^\dagger a+\omega_z \hat{S}_z+y(ae^{i\omega_pt}+a^\dagger e^{-i\omega_pt})\frac{\hat{S}_x}{\sqrt{N}}.
\label{Eq. H}
\end{equation}
Here, $\hat{S}_\alpha$ ($\alpha\in\{x,y,z\}$) are the components of a large spin of length $N/2$, $a$ is the bosonic cavity mode, $y$ is the coupling strength, and $\omega_{0/z/p}$ denote the cavity, atomic, and drive frequency, respectively. This particular realization of a driven Dicke model (with novel time-dependent coupling) was originally introduced in Ref.~\cite{PhysRevLett.104.130401}. The Hamiltonian has been experimentally realized ~\cite{2010Natur.464.1301B,PhysRevLett.107.140402,Klinder3290} and  describes a laser-driven Bose-Einstein condensate coupled to an optical cavity. The time-dependent atom-photon coupling is due to the atom mediating interactions between the driving laser and cavity. Moving to a frame rotating at the drive frequency $\omega_p$, the Hamiltonian takes the form
\begin{equation}
H=\Delta a^\dagger a+\omega_z \hat{S}_z+y(a+a^\dagger)\frac{\hat{S}_x}{\sqrt{N}},
\label{H_spin}
\end{equation}
where $\Delta=\omega_0-\omega_p$ is the cavity detuning. While it is time independent, the Hamiltonian together with the dissipation---to be discussed shortly---describes the dynamics of a non-equilibrium system \cite{PhysRevA.87.023831,2018arXiv180509828K}.  We also remark that there are other experimental realizations of the Dicke model with different microscopic origins. Those include a multilevel atom scheme \cite{Baden14} proposed by Dimer \etal \cite{PhysRevA.75.013804}, as well as driven atoms coupled to a single standing-wave cavity mode \cite{PhysRevLett.91.203001}  proposed by Domokos \etal \cite{PhysRevLett.89.253003} (see Ref.~\cite{2018arXiv180509828K} for a review of these systems).

\begin{table*}
\begin{tabular}{|c|c|c|c|c|c|c|}
\hline 
\multirow{3}*{Observables} &\multirow{3}*{Photon-flux}& \multicolumn{2}{c|}{\multirow{2}*{Correlations}} &  \multicolumn{2}{c|}{\multirow{2}*{Response}} & \multirow{3}*{Finite-size  scaling} \tabularnewline
& &\multicolumn{2}{c|}{}  &  \multicolumn{2}{c|}{} &   \tabularnewline
\cline{3-6}
 & &At criticality &  Away from criticality&  At criticality &  Away from criticality &  \tabularnewline
 \hline
Critical exponent & $\langle a^\dagger a\rangle\propto\delta y^{-\nu} $ & \multicolumn{2}{c|}{\makecell{$iG_\mathrm{ph}^K(t)\equiv\langle \{a(t),a^{\dagger}(0)\}\rangle$ \vspace{.1cm}\\
$\propto\,\, |t|^{-\nu_t}$\vspace{.1cm}}}& \multicolumn{2}{c|}{\makecell{$iG_\mathrm{ph}^R(t)\equiv \Theta(t)\langle [a(t),a^{\dagger}(0)]\rangle$\vspace{.1cm}\\
$\propto \,\,t^{-\nu'_t}$\vspace{.1cm}}} &$\langle a^\dagger a\rangle\propto N^{\alpha}$   \tabularnewline
\hline
\makecell{MB on $\&$ NMB on \\ $T_b=0,~\mu_b\leq0$} & $\begin{cases}
    2-1/s,&s>\frac{1}{2}\\
    0,&s<\frac{1}{2}
  \end{cases}$ & $\begin{cases}
    \text{IR Div.},&s>\frac{1}{2}\\
    1-2s,&s<\frac{1}{2}
  \end{cases}$& $1+s$ &$1-s$&$1+s$&$\begin{cases}
    \frac{2s-1}{3s-1},&s>\frac{1}{2}\\
    0,& s<\frac{1}{2}
  \end{cases}$ \tabularnewline
 \hline
\makecell{MB on $\&$ NMB on \\ $T_b\neq 0$, $\mu_b=0$} & 1 & IR Divergent &$s$ & $1-s$ & $1+s$ & $1/2$ \tabularnewline
\hline
\makecell{MB on $\&$ NMB off \\ $T_b,\mu_b$: NA} & 1 & IR Divergent &  Exp. Decay& IR Divergent & Exp. Decay & $1/2$ \tabularnewline
\hline
\makecell{MB off $\&$ NMB on \\ $T_b=0,~\mu_b\leq0$ } & 0 & $1-s$ & $1+s$ & $1-s$& $1+s$ & $0$ \tabularnewline
\hline
\end{tabular}
\caption{Critical exponents of the photonic field in the different settings described in Table~\ref{situtionTable}. The atomic field exhibits an identical set of critical exponents. Here, we denote Markovian (non-Markovian) bath by MB (NMB) and restrict ourselves to $0<s<1$. When the Markovian bath is present while the non-Markovian bath is absent, the correlation and response functions decay exponentially as $G_{\rm ph}^K(t)\sim e^{-|t| \delta y}/\delta y$ and  $G_{\rm ph}^R\sim \Theta(t)e^{-t \delta y}$, respectively, with $\delta y$ the distance to criticality. The ``IR (infrared) Divergent'' indicates that the corresponding function diverges with system size. The photon-flux and finite-size scaling exponents are the same either when both baths are present with $T_b\neq0$ and $\mu_b=0$ or when the MB (NMB) bath is on (off), and are consistent with those at a thermal phase transition; see the text for the explanation.}
\label{CritEXPTABLE}
\end{table*}

The driven Dicke model possesses a $Z_2$ symmetry as it is invariant under $a\rightarrow -a$ and $\hat{S}_x\rightarrow -\hat{S}_x$. At a sufficiently large coupling strength, the ground state spontaneously breaks the $Z_2$ symmetry and exhibits a large population of the cavity mode with a finite expectation value of the cavity mode $a$. It is convenient to describe spins in terms of bosons via the Holstein-Primakoff transformation,
\begin{equation}
\hat{S}_z=b^\dagger b-N/2,~~~\hat{S}^+=b^\dagger\sqrt{N-b^\dagger b}, ~~~~ \hat{S}^-=(\hat{S}^+)^\dagger,
\end{equation}
where $\hat{S}^{\pm}=\hat{S}^x\pm i \hat{S}^y$.
In the large-$N$ limit, we make an approximation by retaining only the quadratic terms in the Hamiltonian (finite size corrections will be considered in Sec.~\ref{sec:FS}).
We then find
\begin{equation}
H=\Delta a^\dagger a+\omega_z b^\dagger b+\frac{y}{2}(a+a^\dagger)(b^\dagger+b).
\label{Hab}
\end{equation}
At zero temperature, this Hamiltonian exhibits a quantum phase transition at $y_c=\sqrt{\omega_z\Delta}$ \cite{PhysRevA.8.1440,carmichael1973higher,PhysRevA.9.418,HEPP1973360,PhysRevA.7.831,PhysRevE.67.066203,PhysRevA.87.023831,2018arXiv180509828K}. The photon number in the ground state diverges as $\langle a^\dagger a\rangle\sim |y-y_c|^{-\nu}$ as one approaches the critical point from the disordered side; here, $\nu=1/2$ describes the photon-flux exponent. At the critical point, the population diverges in the thermodynamic limit ($N\to \infty$); however, at any finite $N$, it scales as $\langle a^\dagger a\rangle \sim N^{\alpha}$ with $\alpha=1/3$ describing the finite-size scaling exponent \cite{vidal2006finite,PhysRevA.80.023810,PhysRevA.87.023831,2018arXiv180509828K}. We shall see that both photon-flux and finite-size scaling exponents are different when dissipation is included.

\subsection{Dissipation via Schwinger-Keldysh action}\label{SKA_SEC}
We now discuss the effect of dissipation in our system of photons coupled to atoms. 
We consider the usual Markovian dissipation for the cavity mode describing a typical setting in quantum optics, but assume that the atoms are coupled to a sub-ohmic bath that gives rise to non-Markovian dissipation. We shall describe the dynamics due to the drive as well as both Markovian and non-Markovian dissipation. 

We first consider the dissipative dynamics of the cavity mode. This can be properly described by a quantum master equation governing the density matrix of the cavity photons, $\rho$, as
\begin{equation}
\partial_t\rho = -i[\Delta a^\dagger a,\rho]+\kappa (2 a\rho a^\dagger-\{a^\dagger a,\rho\}),
\end{equation}
where $\kappa$ is the decay rate of cavity photons. This master equation has a standard Lindbladian form, owing to the Markovian nature of the dynamics. Next we cast the dissipative dynamics in the form of the Schwinger-Keldysh action. The latter is particularly useful to describe the non-Markovian dissipation of the atomic field. For a review of the Schwinger-Keldysh path-integral formalism, we refer the readers to Refs.~\cite{kamenev2011field,2016RPPh...79i6001S}. 
Within this formalism, the photon operator can be cast in terms of backward and forward complex-valued fields reflecting the evolution of both ket and bra states in the density matrix \cite{kamenev2011field,2016RPPh...79i6001S}. A convenient (Keldysh) rotation of the two fields casts the action in the Keldysh basis. The Keldysh action for the cavity mode is then obtained as
\begin{align}
S_\mathrm{ph}=
\int_\omega(a_{cl}^*,a_{q}^*)\left(\begin{array}{cc}
0 & P^A_\mathrm{ph}(\omega)\\
~P^R_\mathrm{ph}(\omega)& P^K_\mathrm{ph}(\omega)
\end{array}\right)\vect{a_{cl}}{a_{q}},
\end{align}
where $\int_\omega=\int_{-\infty}^\infty\frac{d\omega}{2\pi}$ and $a_{cl/q}$ describe the ``classical/quantum'' fields, respectively. The inverse retarded, advanced, and Keldysh Green's functions of the cavity mode are given by
\begin{equation}
\begin{split}
P^R_\mathrm{ph}(\omega)=(P^A_\mathrm{ph}(\omega))^*&=\omega-\Delta+i\kappa, \\
P^K_\mathrm{ph}(\omega)&=2i\kappa.
\end{split}
\end{equation}
The Green's functions of the cavity mode are then given by $G^{R/A}_{\mathrm{ph}}(\omega)=1/P^{R/A}_{\mathrm{ph}}(\omega)$ and $G^{K}_{\mathrm{ph}}(\omega)=-P^K_\mathrm{ph}(\omega)/(P^{R}_{\mathrm{ph}}(\omega)P^{A}_{\mathrm{ph}}(\omega))$. The above functions satisfy the relation
\begin{equation}
 P^K_\mathrm{ph}(\omega)=P^R_\mathrm{ph}(\omega)F_\mathrm{ph}(\omega)-F_\mathrm{ph}(\omega)P^A_\mathrm{ph}(\omega),
\end{equation}
where the distribution function $F_{\rm ph}(\omega)$ is given by $F_\mathrm{ph}(\omega)=1$. In general, the distribution function reveals whether or not a system is in equilibrium. In thermal equilibrium (for a generic distribution function), we have $F^{\rm Th}(\omega)=\coth(\frac{\omega-\mu}{2T})$ with  $T$ the temperature and $\mu$ the chemical potential, in accordance with the fluctuation-dissipation theorem \cite{kamenev2011field,2016RPPh...79i6001S}. Notice that, at finite temperature but at low frequencies and zero chemical potential (appropriate for photons), the thermal distribution function behaves as $F^{\rm Th}(\omega)\sim 2T/\omega$. At zero temperature, the distribution function becomes $F^{T=0}(\omega)=\sgn(\omega)$. In contrast, the fact that the the cavity-mode distribution function [$F_\mathrm{ph}(\omega)=1$] is symmetric around $\omega=0$ signifies that the system is being probed in the rotating frame \cite{PhysRevA.87.023831}.

We now turn to the dissipation of the atomic mode. Before presenting the action, we first discuss the nature of the non-Markovian bath. The Hamiltonian for the atomic modes coupled to a bosonic bath is given by
\begin{equation}
H=\omega_z b^\dagger b+\sum_k\omega_kc^\dagger_kc^{\phantom{\dagger}}_k+\sum_k (g_k b^\dagger c^{\phantom{\dagger}}_k+g^*_k bc^{\dagger}_k),
\label{BbathHam}
\end{equation}
where $c^\dagger_k$ is the bosonic creation operator for the $k$th bath mode, $g_k$ is the coupling strength for the $k$th bath mode, and $\omega_k$ is the dispersion relation of the bath. 
The bath degrees of freedom can be integrated out leading to an effective description of the bath via the bath spectral density $\rho(\omega)=\sum_k|g_k|^2\delta(\omega-\omega_k)$. 
In particular, the behavior of the spectral density at low frequencies determines the nature of the bath and whether it is ohmic or not \cite{RevModPhys.59.1,Weiss08}. In this work, we shall consider a sub-ohmic bath characterized by the spectral density
\begin{equation}
\rho(\omega)=\gamma\Theta(\omega)\bigg(\frac{\omega}{\omega_z}\bigg)^s\frac{1}{1+(\omega/\omega_M)^2},
\end{equation}
where $0<s<1$ underscores the sub-ohmic nature of the bath. 
Here, $\gamma$ is the dissipation strength and $\omega_M$ is a cutoff frequency; we  take $\omega_M=10^4\omega_b$ throughout this work. 
We shall not discuss the experimental realization of this type of bath, but refer the reader to the proposals and a discussion of their feasibility in Refs.~\cite{PhysRevLett.115.043601,PhysRevA.94.063862,PhysRevA.98.063608}. Finally, by tracing out the bath modes within the standard path-integral formalism \cite{PhysRevLett.115.043601,PhysRevA.94.063862}, one can obtain the Schwinger-Keldysh action for the atomic mode as 
\begin{align}
S_\mathrm{at}=
\int_\omega(b_{cl}^*,b_{q}^*)\bmx
 0 & P^A_{\mathrm{at}}(\omega) \\
P^R_{\mathrm{at}}(\omega) & P^K_\mathrm{at}(\omega)
\emx
\vect{b_{cl}}{b_{q}},
\end{align}
where the inverse Green's functions of the atomic mode are given by
\begin{align}
\label{eqn:eqlabel}
\begin{split}
 P^R_\mathrm{at}(\omega)=(P^A_\mathrm{at}(\omega))^*&=\omega-\omega_z-K^R(\omega),
\\
P^K_\mathrm{at}(\omega)&=2i\pi\rho(\omega),
\end{split}
\end{align}
with the self-energy given by (in the limit $\omega_M\rightarrow\infty$)\footnote{We found that there is a factor of $\pi$ missing from $K^R(\omega)$ when it is numerically evaluated in Refs.~\cite{PhysRevLett.115.043601} and~\cite{PhysRevA.94.063862}. However, this does not affect the numerical value of the critical exponents. We also point out there is a factor of $\pi$ difference between $\rho(\omega)$ in Ref.~\cite{PhysRevLett.115.043601} and that in Ref.~\cite{PhysRevA.94.063862}. To that end, we use the spectral density defined in Ref.~\cite{PhysRevA.94.063862}.} 
\begin{align}
K^{R/A}(\omega)&=\mathcal{P}\int_0^\infty d\omega'\frac{\rho(\omega')}{\omega-\omega'}\mp i\pi\rho(\omega)\nonumber \\
&=\frac{\pi \gamma}{\sin s\pi}\bigg(\frac{|\omega|}{\omega_z}\bigg)^s\bigg(\Theta(\omega)e^{\mp i\pi s}+\Theta(-\omega)\bigg).
\end{align}
The Green's functions of the atomic mode are then given by $G^{R/A}_{\mathrm{at}}(\omega)=1/P^{R/A}_{\mathrm{at}}(\omega)$ and $G^{K}_{\mathrm{at}}(\omega)=-P^K_\mathrm{at}(\omega)/(P^{R}_{\mathrm{at}}(\omega)P^{A}_{\mathrm{at}}(\omega))$. The inverse Green's functions of the atomic mode obey the following fluctuation-dissipation relation,
\begin{equation}
 P^K_\mathrm{at}(\omega)=P^R_\mathrm{at}(\omega)F_\mathrm{at}(\omega)-F_\mathrm{at}(\omega)P^A_\mathrm{at}(\omega),
\end{equation}
where $F_\mathrm{at}(\omega)=\sgn(\omega)$. This indicates that the uncoupled atomic mode is in equilibrium at zero temperature. Finite temperature can be included by substituting $P^K_\mathrm{at}(\omega)\rightarrow P^K_\mathrm{at}(\omega)\coth(\frac{\omega -\mu_b}{2T_b})$ \cite{PhysRevA.94.063862}, where $\mu_b\leq0$ and $T_b$ define the chemical potential and temperature of the non-Markovian bath (in units where Boltzmann's constant, $k_B$, is unity). This gives $F_\mathrm{at}(\omega)=\coth(\frac{\omega -\mu_b}{2T_b})$, indicating that the (uncoupled) atomic mode is in thermal equilibrium. The cavity mode is still assumed at zero temperature since the coupling to the Markovian bath---at optical frequencies---is unaffected by the finite temperature of the non-Markovian bath \cite{PhysRevA.94.063862}.

Finally, the contribution of the coupling term in the Hamiltonian, Eq.~\eqref{Hab}, to the action is given by the Keldysh action (in the time-domain),
\begin{align}
S_{\rm ph-at}=-\frac{y}{2}\int_t[(a_q+a_q^*)(b_{cl}+b_{cl}^*)+(a_{cl}+a_{cl}^*)(b_{q}+b_{q}^*)],
\label{sab}
\end{align}
where $\int_t=\int_{-\infty}^{\infty}dt$ and all fields are evaluated at the same time. 
The total action describing the full driven-dissipative dynamics of our model is then given by
\begin{equation}
\label{FULLACT}
S=S_{\rm ph}+S_{\rm at}+S_{\rm ph-at}.
\end{equation}
(The nonlinear terms required to characterize finite-size effects are discussed in Sec.~\ref{sec:FS}.) This driven-dissipative system undergoes a superradiant phase transition at the critical point given by $y_c=\sqrt{\frac{\Delta^2+\kappa^2}{\Delta}\omega_z}$ \cite{PhysRevA.84.043637,PhysRevA.87.023831,2018arXiv180509828K,PhysRevLett.115.043601,PhysRevA.94.063862}. One interesting point to note  is that the position of the critical point, $y_c$, does not depend on the colored bath. This is because the critical point is determined exactly via a mean-field analysis of the action at $\omega=0$, at which point the contribution of the non-Markovian bath vanishes. 
In this work, we shall restrict ourselves to the disordered side, $y\leq y_c$. But we note that the critical exponent characterizing ordering is not sensitive to temporal fluctuations and is thus purely determined by the mean-field analysis at $\omega=0$; see Ref.~\cite{PhysRevA.87.023831} for a detailed treatment of critical exponents in the ordered phase. In addition, we restrict ourselves to sub-ohmic colored baths ($s<1$). For $s\ge 1$ and $\kappa\neq0$, the critical behavior is identical to that in the absence of non-Markovian bath consistent with the results reported by Nagy and Domokos~\cite{PhysRevLett.115.043601,PhysRevA.94.063862}.

\section{Low-Frequency Field Theory}\label{sec:LEFT}

In this section, we explicitly derive the low-frequency description of our model [defined in Eq.~\eqref{FULLACT}] and further discuss the nature of its non-equilibrium phase transition. The key observation is that the order parameter corresponding to the $Z_2$ symmetry is of the Ising type, and thus an effective field-theoretical description in terms of a single scalar (real) field should exist. On the other hand at the microscopic level, both atomic and cavity fields are characterized by four complex-valued fields (eight real-valued fields). The key technical step (after integrating out the atomic field) is a non-trivial rotation of photonic fields [see Eq.~\eqref{transformafields} and Fig.~\ref{rotation_fig}], which gives a set of four real-valued fields, two of which are always gapped and can be safely integrated out without generating long-range coupling in time.The dynamics of the remaining two fields can be then turned into a stochastic equation that involves a single real-valued field describing the order parameter. This makes the calculation of various low-frequency properties analytically tractable. The low-frequency properties of this model (and other closely related models in Sec.~\ref{sec:othermodels}) are summarized in Table~\ref{situtionTable}.

\subsection{Field theory of the cavity mode}
In this section, we focus only on the low-frequency theory of the cavity field. 
The end result is the Schwinger-Keldysh action in terms of the classical and quantum components of a single (scalar) real field---mimicking the Ising nature of the order parameter---which captures the critical properties of the model. The reader who is not interested in the technical steps of this derivation may skip directly to  Sec.~\ref{MAINSECND}. In Appendix~\ref{sec:LEATOMICFIELD}, we present the analogous steps to derive the low-frequency theory of the atomic field, which is 
shown to be of the same form as that of the cavity field, albeit with different coefficients. Therefore, both cavity and atomic modes exhibit the same critical behavior and exponents. 

We start by integrating out the atomic degrees of freedom in Eq.~\eqref{FULLACT} to find the effective action\footnote{Here, we have absorbed a factor of 1/2 into the $a$ fields to match the notation of Refs.~\cite{PhysRevLett.115.043601,PhysRevA.94.063862}. This does not effect the critical properties.} for the cavity mode \cite{PhysRevLett.115.043601,PhysRevA.94.063862},
\begin{equation}
S_\mathrm{ph}^{\mathrm{eff}} =\int_\omega
\b{v}_{\mathrm{ph}}^\dagger 
\bmx
 0 & \b{P}^A_{\mathrm{ph}}(\omega) \\
\b{P}^R_{\mathrm{ph}}(\omega) & \b{P}^K_{\mathrm{ph}}(\omega)
\emx
 \b{v}^{\phantom{\dagger}}_{\mathrm{ph}}\,,
 \label{eq:S_eff_a}
\end{equation}
where 
\begin{equation}
\b{v}_\mathrm{ph}^\dagger(\omega)=(a_{cl}^*(\omega),a_{cl}(-\omega),a_{q}^*(\omega),a_{q}(-\omega)),
\end{equation}
and $\b{P}^R_{\mathrm{ph}}(\omega)$, $\b{P}^A_{\mathrm{ph}}(\omega)$, and $\b{P}^K_{\mathrm{ph}}(\omega)$ are $2 \times 2$ matrices. The inverse retarded Green's function, $\b{P}^R_{\mathrm{ph}}(\omega)$, and inverse advanced Green's function, $\b{P}^A_{\mathrm{ph}}(\omega)$, are given by
\begin{align}
\b{P}^R_{\mathrm{ph}}&(\omega)=(\b{P}^A_{\rm ph}(\omega)^{-1})^\dagger=\nonumber \\
&\bmx
 P^R_\mathrm{ph}(\omega)+\Sigma_\mathrm{ph}^R(\omega) &\Sigma_\mathrm{ph}^R(\omega)   \\
\Sigma_\mathrm{ph}^R(\omega)  & P^A_\mathrm{ph}(-\omega)+\Sigma_\mathrm{ph}^R(\omega)
\emx,
\label{bP for photons}
\end{align}
where $\Sigma_\mathrm{ph}^R(\omega)$ is the photon self-energy and is given by
\begin{equation}
\Sigma_\mathrm{ph}^R(\omega)=-\frac{y^2}{4}\bigg(\frac{1}{P^R_\mathrm{at}(\omega)}+\frac{1}{P^A_\mathrm{at}(-\omega)}\bigg).
\end{equation}
The inverse Keldysh Green's function, describing the effective bath to which the cavity photons are coupled, is given by
\begin{align}
\b{P}^K_{\mathrm{ph}}(\omega)=
\bmx
P^K_\mathrm{ph}(\omega)+d(\omega) &  d(\omega)  \\
d(\omega)  & P^K_\mathrm{ph}(\omega)+d(\omega)
\emx,\end{align}
with
\begin{equation}
d(\omega)=\frac{y^2}{4}\bigg(\frac{P^K_\mathrm{at}(\omega)}{|P^R_\mathrm{at}(\omega)|^2}+\frac{P^K_\mathrm{at}(-\omega)}{|P^R_\mathrm{at}(-\omega)|^2}\bigg).
\end{equation}

\begin{figure}[t]
  \centering\includegraphics[width=.45\textwidth]{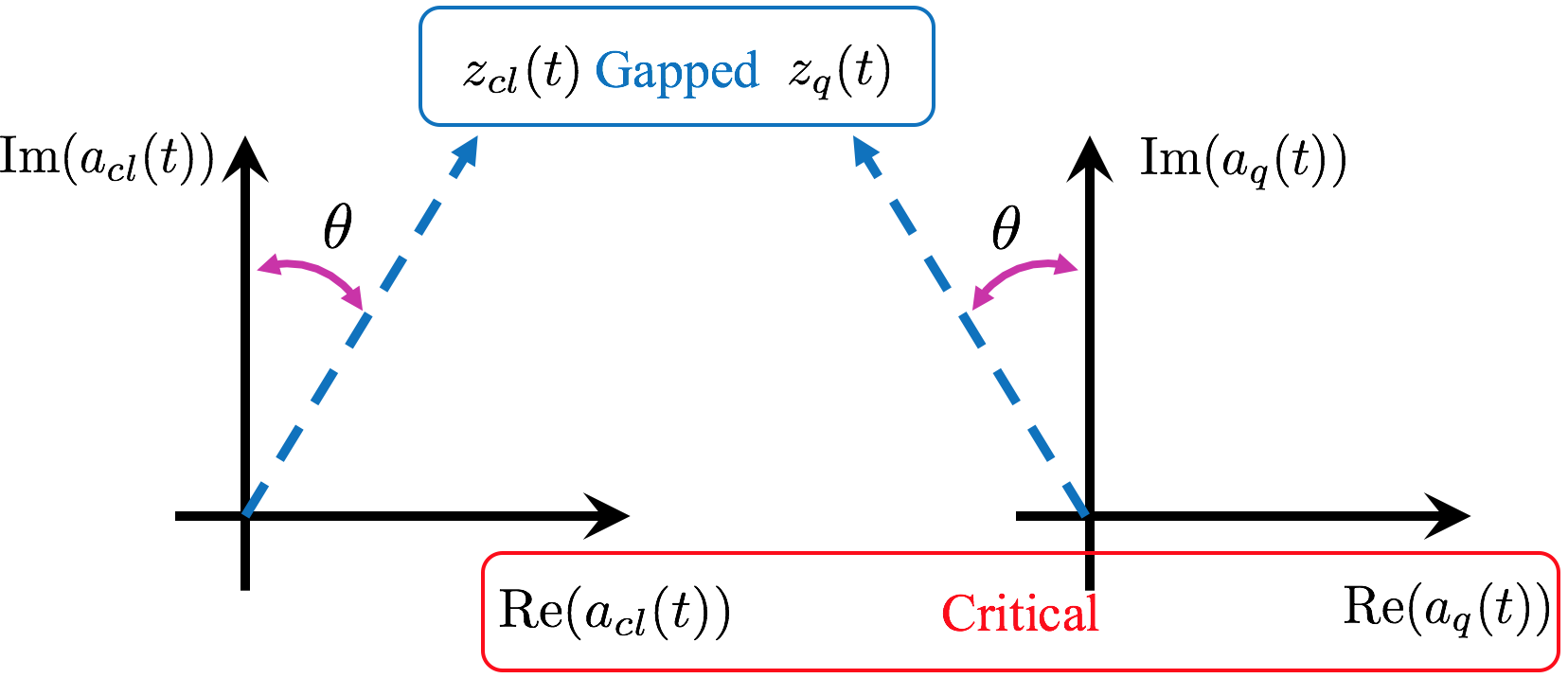}
  \caption{(color online) Schematic illustration of the non-trivial field rotation [see Eq.~\eqref{transformafields}]; we have defined the angle $\theta=\pi/2-|p|$ with $p=\arctan(-\Delta/\kappa)$. The dashed line defined by this angle corresponds to a gapped field while the real parts of the fields $a_{cl/q}$ capture the critical behavior.}
  \label{rotation_fig}
\end{figure}

The goal of this section is to simplify Eq.~\eqref{eq:S_eff_a} so we can understand its low-frequency properties analytically. To this end, we change our basis to two real fields. At the first glance, this might not seem like a useful transformation, but we shall see that one of the two fields is always gapped and can thus be integrated out without generating long-range coupling in time. We note a similar change of basis was carried out in Ref.~\cite{PhysRevB.93.014307}; see also Ref.~\cite{PhysRevA.95.043826}. More specifically, we seek a change of basis where $P^R_{\mathrm{ph}}(\omega)$ is diagonal at zero frequency ($\omega=0$) and at the critical point ($y=y_c$). To proceed, we write $\b{P}^R_{\mathrm{ph}}(\omega)$ as
\begin{align}
\b{P}^R_{\mathrm{ph}}(\omega,y)=\b{P}^R_{\mathrm{ph}}(0,y_c)+
\bmx
\delta\Sigma_\mathrm{ph}^R+\omega  & \delta\Sigma_\mathrm{ph}^R  \\
\delta\Sigma_\mathrm{ph}^R  & \delta\Sigma_\mathrm{ph}^R-\omega
\emx,
\end{align}
where $\delta\Sigma_\mathrm{ph}^R=\Sigma_\mathrm{ph}^R(\omega,y)-\Sigma_\mathrm{ph}^R(0,y_c)$; here, we have explicitly included the dependence of $\b{P}^R_{\mathrm{ph}}(\omega)$ and $\Sigma_\mathrm{ph}^R(\omega)$ on $y$. We stress that this expression is exact and no approximations have been made yet [given Eq.~\eqref{eq:S_eff_a} as a starting point]. We define a set of new classical and quantum fields as
\begin{align}
\left(\begin{array}{cc}
x_{cl/q}(\omega) \\
 z^*_{cl/q}(-\omega) \end{array}\right)=\b{R}_{\mathrm{ph},cl/q}\left(\begin{array}{cc}a_{cl/q}(\omega) \\
 a^*_{cl/q}(-\omega) \end{array}\right),
\end{align}
where $\b{R}_{\mathrm{ph},cl/q}$ are $2 \times 2$ matrices. One can easily see that choosing
\begin{align}
\b{R}_{\mathrm{ph},cl/q}=\left(\begin{array}{cc}
 1  & 1  \\
 \pm e^{\mp ip}  &  \pm e^{\pm ip} 
\end{array}\right),
\label{transformafields}
\end{align}
where $p=\arctan(-\Delta/\kappa)$, diagonalizes $\b{P}^R_\mathrm{ph}(0,y_c)$ and ensures that the new fields are real in the time domain\footnote{In the frequency domain, they satisfy $x^*(\omega)=x(-\omega)$ and similarly for the $z$ field.}; see Fig.~\ref{rotation_fig}. The effective action is then given by
\begin{equation}
S_{\mathrm{ph}}^{\mathrm{eff}} = \int_\omega
\b{\tilde{v}}_{\mathrm{ph}}^\dagger 
\bmx
 0 & \b{\tilde{P}}^A_{\mathrm{ph}}(\omega) \\
\b{\tilde{P}}^R_{\mathrm{ph}}(\omega) & \b{\tilde{P}}^K_{\mathrm{ph}}(\omega)
\emx
 \b{\tilde{v}}^{\phantom{\dagger}}_{\mathrm{ph}}\,,
\end{equation}
in the new basis defined as 
\begin{equation}
\b{\tilde{v}}_{\mathrm{ph}}^\dagger(\omega)=(x_{cl}^*(\omega),z^*_{cl}(\omega),x_{q}^*(\omega),z^*_{q}(\omega)).
\end{equation}
The inverse retarded Green's function in the rotated basis is given by
\begin{align}
\b{\tilde{P}}^R_{\mathrm{ph}}(\omega)=
\bmx
\delta\Sigma_\mathrm{ph}+i\omega \chi & -\frac{i}{2}\omega\sqrt{1+\chi^2}  \\
 \frac{i}{2}\omega\sqrt{1+\chi^2}  & -M
\emx,
\end{align}
where $\chi\equiv\frac{\kappa}{\Delta}$ and $M\equiv\frac{\Delta}{2}(1+\chi^2)$ defines the ``gap'' of the $z$ field. Notice that, at zero frequency, the matrix becomes diagonal. The inverse Keldysh Green's function in the rotated basis is given by
\begin{align}
{\b{ \tilde{P}}}^K_{\mathrm{ph}}(\omega)=
\bmx
d(\omega)+2i\chi M&   i\kappa\chi\sqrt{2M/\Delta} \\
 i\kappa\chi\sqrt{2M/\Delta}  & 2i\chi M
\emx.
\label{DKnewbasisafield}
\end{align}

We now make the crucial observation that both the classical and quantum $z$ fields are always gapped. In other words, the correlations involving $z$ remain finite and are insensitive to criticality. We can then safely integrate them out to obtain a low-frequency description of the model in terms of the $x$ fields. This particularly makes sense as the field $x_{cl}$ corresponds to the expectation value $\langle a+a^\dagger\rangle$ which is nothing but the order parameter near the superradiant phase transition.
Integrating out the gapped fields ($z_{cl}$ and $z_q$), we obtain an effective action as 
\begin{align}
S_{x}^{\rm eff}=
\int_\omega(x_{cl}^*,x^*_{q})
\bmx
0 &  P^A_{x}(\omega) \\
P^R_{x}(\omega) & P_x^{K}(\omega)
\emx
\left(\begin{array}{cc}
x_{cl} \\
x_{q}
\end{array}\right),
\label{c_aaction}
\end{align}
where
\begin{align}\label{Eq. PR_x}
P^R_{x}(\omega)=(P^A_{x}(\omega))^*=
\delta\Sigma^R_\mathrm{ph}+i\omega\chi-\frac{\omega^2}{2\Delta},
\end{align}
and 
\begin{align}
P_x^{K}(\omega)=[\tilde{\b{P}}^K_{\mathrm{ph}}]_{11}+\frac{\omega^2}{4M^2}(1+\chi^2)[\tilde{\b{P}}^K_{\mathrm{ph}}]_{22}.
\label{exactKc}
\end{align}
The matrix elements of ${\b{ \tilde{P}}}^K_{\mathrm{ph}}(\omega)$ are given in Eq.~\eqref{DKnewbasisafield}. We remind the reader that these expressions are still exact.

\subsection{Low-frequency limit}\label{MAINSECND}

We now take the low-frequency limit (small $\omega$) near the critical point ($y_c-y\equiv \delta y\ll y_c$) of Eq.~(\ref{c_aaction}) with both the Markovian and non-Markovian baths present; in this section, we assume that $T_b,\mu_b=0$. We note that for $\mu_b<0$ and $T_b=0$, the inverse Keldysh Green's function is given by $P^K_\mathrm{at}(\omega)\sgn(\omega-\mu_b)$. Since $P^K_\mathrm{at}(\omega)$ is only nonzero for $\omega>0$, we have  $P^K_\mathrm{at}(\omega)\sgn(\omega-\mu_b)=P^K_\mathrm{at}(\omega)$. Thus, the low-frequency theory at zero temperature is independent of the chemical potential (which is always equal to or less than zero for bosonic systems) of the non-Markovian bath. In Sec.~\ref{finiteTEMPTHEORY}, we show that the same low-frequency field theory emerges when $T_b,\mu_b\neq0$; however, a different behavior ensues when $T_b\neq 0, \mu_b=0$, i.e., when the non-Markovian bath is at a finite temperature with a vanishing chemical potential. 

Finally, we make an approximation by expanding various terms near the critical point and at small frequencies. From Eq.~\eqref{Eq. PR_x}, the inverse retarded Green's function of the $x$ field at low frequencies is given by
\begin{align}
P^R_{x}(\omega)\approx
 -r+
\bigg|\frac{\omega}{\omega_z}\bigg|^s\bigg(iv_I\sgn(\omega)-v_R\bigg),
\label{grcase3}
\end{align}
where $r=y_c\delta y/\omega_z$ is proportional to the distance from criticality, $v_I=\frac{y_c^2}{4\omega_B^2}\pi\gamma$ and $v_R=v_I\cot(\pi s/2)$. From Eq.~\eqref{exactKc}, the inverse Keldysh Green's function in the low-frequency limit is given by
\begin{align}
P^K_{x}(\omega)\approx i\kappa(1+\chi^2)\equiv i2\kappa_{\rm eff},
\label{DKcase1}
\end{align}
where, in the last equality, we have defined $\kappa_{\rm eff}\equiv\kappa(1+\chi^2)/2=\kappa(1+\kappa^2/\Delta^2)/2$. 
Interestingly, the effective bath for the low-frequency theory is Markovian, however, the low-frequency retarded Green's function is significantly modified by the non-Markovian bath. This leads to the distribution function,
\begin{align}
F_{x}(\omega)=\frac{G^K_{x}(\omega)}{G^R_{x}(\omega)-G^A_{x}(\omega)}=
\frac{\kappa_{\rm eff}}{v_I}\bigg|\frac{\omega_z}{\omega}\bigg|^s\sgn(\omega).
\end{align}
This distribution function clearly indicates that the system is not in thermal equilibrium (as it is neither of the form $\sgn(\omega)$ nor is it $2T_{\rm eff}/\omega$ for some constant $T_{\rm eff}$), hence the corresponding phase transition is not thermal and is genuinely non-equilibrium. Nevertheless, we can provide an intriguing interpretation of the above equation by introducing a frequency-dependent effective temperature via $F_x(\omega)\sim 2T_{\rm eff}(\omega)/\omega$. This yields an effective temperature that scales as $T_{\rm eff}(\omega)\sim|\omega|^{1-s}$ and vanishes in the limit $\omega \to 0$. This fact might suggest that a quantum critical behavior (typically arising at or close to zero temperature) could emerge in this system. 
In spite of this expectation, we shall argue in the next (sub)section that the phase transition is classical in nature.

It is helpful to write the inverse Green's function in Eq.~\eqref{grcase3} as $-r-v(-i\omega)^s$ where $v={v_I}/[\sin(\frac{\pi s}{2})\omega_z^s]$. This already indicates that the structure of the Green's function in the complex plane is given by branch cuts rather than poles. Also it makes the causal structure of the Green's function clear\footnote{The retarded (advanced) Green's functions in the frequency domain can be analytically continued to the entire complex plane with a branch cut from the origin to $- i\infty$ ($+ i\infty$). The corresponding Green's function 
is causal (anti-causal) in the time domain since the integration contour can be closed in the upper (lower) half plane for $t<0$ ($t>0$).}. Furthermore, it allows us to make an intriguing connection to fractional derivatives. The central objects in this context are Liouville fractional derivatives $D_{\pm}^s$ whose Fourier transform are given by $(\mp i\omega)^s$ \cite{samko1993fractional,kilbas2006theory}; these are precisely what we have encountered in the inverse Green's functions. Alternatively, one can use the integral definition of $D_{\pm}^s$ in the time domain as
\begin{align}
\begin{split}
D_{+}^sf(t)&=\frac{1}{\Gamma(1-s)}\frac{d}{dt}\int_{-\infty}^t\frac{f(t')}{(t-t')^s}dt', \\
D_{-}^sf(t)&=\frac{1}{\Gamma(1-s)}\frac{d}{dt}\int_{t}^\infty\frac{f(t')}{(t'-t)^s}dt'.
\end{split}
\end{align}
Using fractional derivatives allows one to write the low-frequency action in the temporal domain as
\begin{align}
S_{x}^{\rm eff}=\int_t x_q (-v \partial_t^s-r)x_{cl}+
2i\kappa_{\rm eff}x_q^2.
\label{caseIrealtimeaction}
\end{align}
Here, we have defined $D_+^s\equiv\partial_t^s$ for notational convenience. This yields the Langevin equation
\begin{equation}
 (v \partial_t^s+r)x_{cl}(t)=\xi(t),
\end{equation}
with the noise $\xi(t)$ correlated as
\begin{equation}
    \langle \xi(t)\xi(t')\rangle=2\kappa_{\rm eff}\delta(t-t'),
\end{equation}
hence, the white noise as opposed to the colored noise. Despite the short-range (in time) correlations of the noise, the equation of motion involves a fractional derivative with long-range tails in time. Therefore, in sharp contrast with fractional Langevin equations that describe the dynamics under equilibrium conditions \cite{Lutz01}, the fluctuation-dissipation theorem is strongly violated here. As a result, the system would not equilibrate to an (effectively) thermal state at long times. Alternatively, one can consider the Fokker-Planck equation for the probability distribution of the photonic field. For $s>1/2$, we conjecture the fractional Fokker-Planck equation
 \begin{align}
     \partial_tP(x,t)=
     Ar\partial_t^{1-s}\partial_x(xP)+B\partial_t^{2(1-s)}(\partial_x^2P).
\label{FPEQ}
 \end{align}
Here, $P=P(x,t)$ is the probability distribution function of the photonic field (or the atomic field, as they are described by the same low-frequency theory), and $A$ and $B$ are phenomenological coefficients. Using simple scaling arguments, we show in Appendix~\ref{sec:FPEQ} that this equation correctly reproduces the critical exponents derived explicitly in the next section. We should contrast this equation with the conventional form of the fractional Fokker-Planck equation \cite{Metzler99}:
 \begin{align}
     \partial_tP(x,t)=
     \partial_t^{1-\delta}\left[\frac{1}{\eta}\partial_x(V'(x) P)+D\partial_x^2P\right].
\label{FPEQ-standard}
 \end{align}
Here, the exponent $\delta$ characterizes the anomalous diffusion,
$V(x)\sim r x^2$ is the external potential ($V'=dV/dx$), $\eta$ is a friction coefficient and $D$ is a diffusion constant. At long times (and for $r\neq 0$), the steady state is determined by setting the expression in the bracket to zero that is nothing by the standard Fokker-Planck operator, which thus ensures equilibrium. In the absence of the external potential, \cref{FPEQ,FPEQ-standard} coincide by identifying $\delta=1-2s$. However, in the presence of a confining potential, our conjectured fractional Fokker-Planck equation \eqref{FPEQ} involves two distinct fractional derivatives with different exponents, and thus does not guarantee thermal equilibrium even at late times. This is yet another manifestation of the genuinely non-equilibrium nature of the dynamics.

\subsection{Nature of the non-equilibrium phase transition}\label{MAINSECND}

To determine the (classical or quantum) nature of the phase transition, we first determine the scaling dimensions of both the classical and quantum fields. Following a scaling analysis similar to that in Ref.~\cite{PhysRevA.87.023831}, it is straightforward to see that, at the critical point, the action [Eq.~\eqref{caseIrealtimeaction}] is invariant under the scaling transformation
\begin{align}
\begin{split}
t\rightarrow &\,\lambda t, \\
x_{cl}(t)\rightarrow \lambda^{s-1/2} x_{cl}(t),\qquad 
& x_{q}(t)\rightarrow\frac{1}{\sqrt{\lambda}}x_{q}(t).
\end{split}
\label{case3scaling}
\end{align}
Therefore, the classical and quantum fields have different scaling dimensions; specifically, the scaling dimension of the quantum field is more negative than that of the classical field. It is interesting to note that the relative scaling of classical and quantum fields is set by the scaling of the frequency-dependent effective temperature introduced in the previous (sub)section. A quantum critical behavior requires the same scaling dimensions for the classical and quantum fields, which in turn requires not only the (effective) temperature to vanish as $\omega \to 0$ but to vanish sufficiently fast, at least linearly with $\omega$. However, in our model (with $0<s<1$), ``quantum vertices'' are less relevant than their classical counterparts.

To be more specific, we should derive the interaction terms in our model. The Hamiltonian can be expanded to first order in $1/N$ in terms of Holstein-Primakoff bosons as 
\begin{equation}
H_{\rm int}=-\frac{y}{4N}(a+a^\dagger)b^\dagger(b^\dagger+b)b.
\end{equation}
The corresponding term in the action is simply given by taking the (adjoint) operators to (complex-conjugated) fields on the forward branch and subtracting a similar term in the backward branch. Recasting the action in the Keldysh basis in terms of classical and quantum fields, we find (see also Ref.~\cite{PhysRevA.87.023831})
\begin{align}
S_{\rm int}=\frac{y}{4N}\int_t \,\, &\Big[(a_{cl}+a^*_{cl})(b_{q}+b^*_{q})\nonumber \\ &+(a_{q}+a^*_{q})(b_{cl}+b^*_{cl})\Big](b^*_{cl}b_{cl}+b^*_{q}b_{q})\nonumber \\
+&\Big[(a_{cl}+a^*_{cl})(b_{cl}+b^*_{cl})+\nonumber \\ &+(a_{q}+a^*_{q})(b_{q}+b^*_{q})\Big](b^*_{cl}b_{q}+b^*_{q}b_{cl}).
\label{sint}
\end{align}
 Here, we have absorbed a factor of 1/2 into both $a$ and $b$ fields to match the notation of Refs.~\cite{PhysRevLett.115.043601,PhysRevA.94.063862}. The total action should be then redefined to include the interaction term together with the photonic and atomic actions as well as their bilinear coupling, $S_{\rm ph-at}$ [Eq.~\eqref{sab}], that is, $S \to S=S_{\rm at}+S_{\rm ph}+S_{\rm ph-at}+S_{\rm int}$. Our goal is to obtain an effective action for the $x$ fields---characterizing the order parameter---including the finite-$N$ corrections. To that end, we integrate out the atomic field and express the resulting effective action in the non-trivial basis just introduced. We then integrate out the gapped ($z$) fields and obtain the effective action for the $x$ fields. These two technical steps are presented in detail in Appendix~\ref{sec:INTERACTIONSAEFF}. The resulting contribution to the effective action for the $x$ fields is given by 
\begin{equation}
S^{\rm eff}_{{\rm int},x}=-\frac{g_{\rm ph}}{N}\int_t x_q(t)x_{cl}^3(t)+\dots,
\label{EFFECTIVEACTIONA}
\end{equation}
where $g_{\mathrm{ph}}=\frac{(\Delta^2+\kappa^2)^2}{2\Delta^2\omega_z}$. Here the ellipsis represent terms that are less relevant in the renormalization-group sense; for instance, this includes the term $x_{q}^3 x_{cl}$, which, due to the scaling dimension of the quantum field, would be less relevant and can be dropped. The value of $g_{\mathrm{ph}}$ is in agreement with the coefficient obtained in Ref.~\cite{PhysRevA.87.023831} by investigating the mean-field equations of the system (upon taking into account the slight difference in field definitions); see Appendix \ref{sec:INTERACTIONSMEANFIELD}. With the quantum field appearing at most quadratically, the action can be then converted into a Langevin equation that is given by
\begin{equation}
 (v \partial_t^s+r)x_{cl}(t)+\frac{g_{\mathrm{ph}}}{N} x^3_{cl}(t)=\xi(t), 
 \label{FULLLANG}
\end{equation}
with the noise correlations $\langle \xi(t)\xi(t')\rangle=2\kappa_{\rm eff}\delta(t-t')$. Therefore, the dynamics of the system is governed by a classical stochastic equation---albeit with long-range tails in time---and thus the non-equilibrium phase transition in Ref.~\cite{PhysRevLett.115.043601} is classical (and not quantum) in nature. This equation represents one of the main results of this work. 

\subsection{Finite temperature}\label{finiteTEMPTHEORY}
We now consider finite temperature and chemical potential (of the colored bath). For bosons, the chemical potential is always less than or equal to zero \cite{griffiths2016introduction}. 
At finite temperature and/or chemical potential, the inverse retarded and advanced Green's functions are unchanged and are given by Eq.~\eqref{grcase3}, while the inverse Keldysh Green's function should be modified. This is simply because finite temperature changes the distribution function while the response functions (in the absence of nonlinear interactions) are unaffected. 
Our main observation is that, at small frequencies and finite chemical potential and temperature, the Keldysh action changes as $P^K_\mathrm{at}(\omega)\to P^K_\mathrm{at}(\omega)\coth(\frac{\omega -\mu_b}{2T_b})\approx P^K_\mathrm{at}(\omega)\coth(-\mu_b/2T_b)$, where $P^K_\mathrm{at}(\omega)$ denotes the distribution function at zero temperature and chemical potential. However, $P^K_\mathrm{at}(\omega)$ vanishes with the frequency as $\omega\to 0$ just in the same way, and thus there is no change to the low-frequency theory presented before. This is not in agreement with the results presented in Ref.~\cite{PhysRevA.94.063862}, where the critical exponent was found to change as a function of the chemical potential of the colored bath at finite temperature. In Sec.~\ref{FiniteTempEXP}, we explicitly show that this is due to the fact that the authors did not consider distances close enough to criticality, i.e., sufficiently small $\delta y/y_c$.

Next we turn to the case when the temperature is finite while the chemical potential is zero, i.e., when $T_b\neq 0$ and $\mu_b=0$. We shall see that the low-frequency theory is significantly modified which also leads to different critical exponents. In this case too, the inverse retarded Green's function is unchanged; however, the inverse Keldysh Green's function should be modified to 
\begin{equation}\label{DK finite T}
P^K_{x}(\omega) \to P^K_{x}(\omega)\coth\left(\frac{\omega}{2T_b}\right) 
\approx i\frac{4v_IT_b}{\omega_z^{s}}\frac{1}{|\omega|^{1-s}}.
\end{equation}
The corresponding distribution function is $F_{x}(\omega)=2T_b/\omega$, indicating that the system is in effective thermal equilibrium at temperature $T_b$. Nevertheless, we stress that this effective behavior only holds at low frequencies while the distribution function is a rather complicated function of frequency away from this limit [distinct from the characteristic thermal distribution $\coth(\omega/2T_b)$]; see Eq.~\eqref{exactKc}. The appearance of an effective temperature even in settings far from equilibrium is well known and occurs in a wide variety of physical systems (see for example, Refs.~\cite{RevModPhys.65.851,PhysRevA.87.023831,2016RPPh...79i6001S}). 
It also turns out that the effective temperature of the system is the same as the temperature of the non-Markovian bath; this should be attributed to the fact that the low-frequency population is dominated by the non-Markovian bath and even diverges at $\omega\to 0$ as one can see from Eq.~\eqref{DK finite T}.

Next, we write the action in the time domain:
 \begin{align}
S_{x}^{\rm eff}=& \int_{t} x_q(t) (-v \partial_t^s-r)x_{cl}(t)\nonumber \\ &-ia_svT_b\int_{t}\int_{t'} \frac{x_{q}(t)x_{q}(t')}{ |t-t'|^{s}},
\label{c_aaction_case1}
\end{align}
with $a_s=(2/\pi)\sin^2(\frac{\pi s}{2})\Gamma(s)$. The low-frequency field theory is clearly different from the same model at $T_b,\mu_b=0$; cf. Eq.~\eqref{caseIrealtimeaction}. Indeed, the new action is invariant at the critical point under the scaling transformation
\begin{align}
\begin{split}
t\rightarrow &\,\lambda t, \\
x_{cl}(t)\rightarrow \lambda^{s/2} x_{cl}(t),~\quad &
x_{q}(t)\rightarrow \lambda^{s/2-1}x_{q}(t).
\end{split}
\end{align}
In this case too, the scaling dimension of the quantum field is more negative than that of the classical field, rendering the phase transition classical in nature. However, in contrast with the previous case where $T_b,\mu_b=0$, the transition at $\mu_b= 0$ and $T_b\ne 0$ is effectively thermal as an effective temperature ($T_{\rm eff}=T_b$) emerges and equilibrium fluctuation-dissipation relations can be established at low frequencies \cite{Lutz01}. More generally, the latter (the limit where $\mu_b= 0$ and $T_b\ne 0$) is also described by a Langevin equation similar to Eq.~\eqref{FULLLANG} but with the important difference that the noise has long-range temporal correlations; in other words, the noise is colored rather than white. 
In the next section, we show that this model leads to a different set of critical exponents.

\section{Critical Exponents}\label{sec:CORR}

In this section, we analytically calculate various critical exponents using the low-frequency theory developed in the previous section. These results are presented in Table~\ref{CritEXPTABLE}, along with the results for several closely related models. We then compare them to the exponents obtained from numerical integration of the exact correlation functions. We find excellent agreement between the two methods and the previous numerical results of Nagy and Domokos, except at finite temperature (see Sec.~\ref{FiniteTempEXP}). We resolve this difference by showing that one should consider close enough distances to criticality (see Fig.~\ref{temp}). We note that the atomic field acquires the same critical exponents as the photonic field since they are described by the same low-frequency theory up to multiplicative factors. This is explicitly shown in Appendix~\ref{sec:LEATOMICFIELD}.

\subsection{Photon-flux exponent}\label{PFEXP}

\begin{figure}
  \centering\includegraphics[width=.43\textwidth, height=6cm]{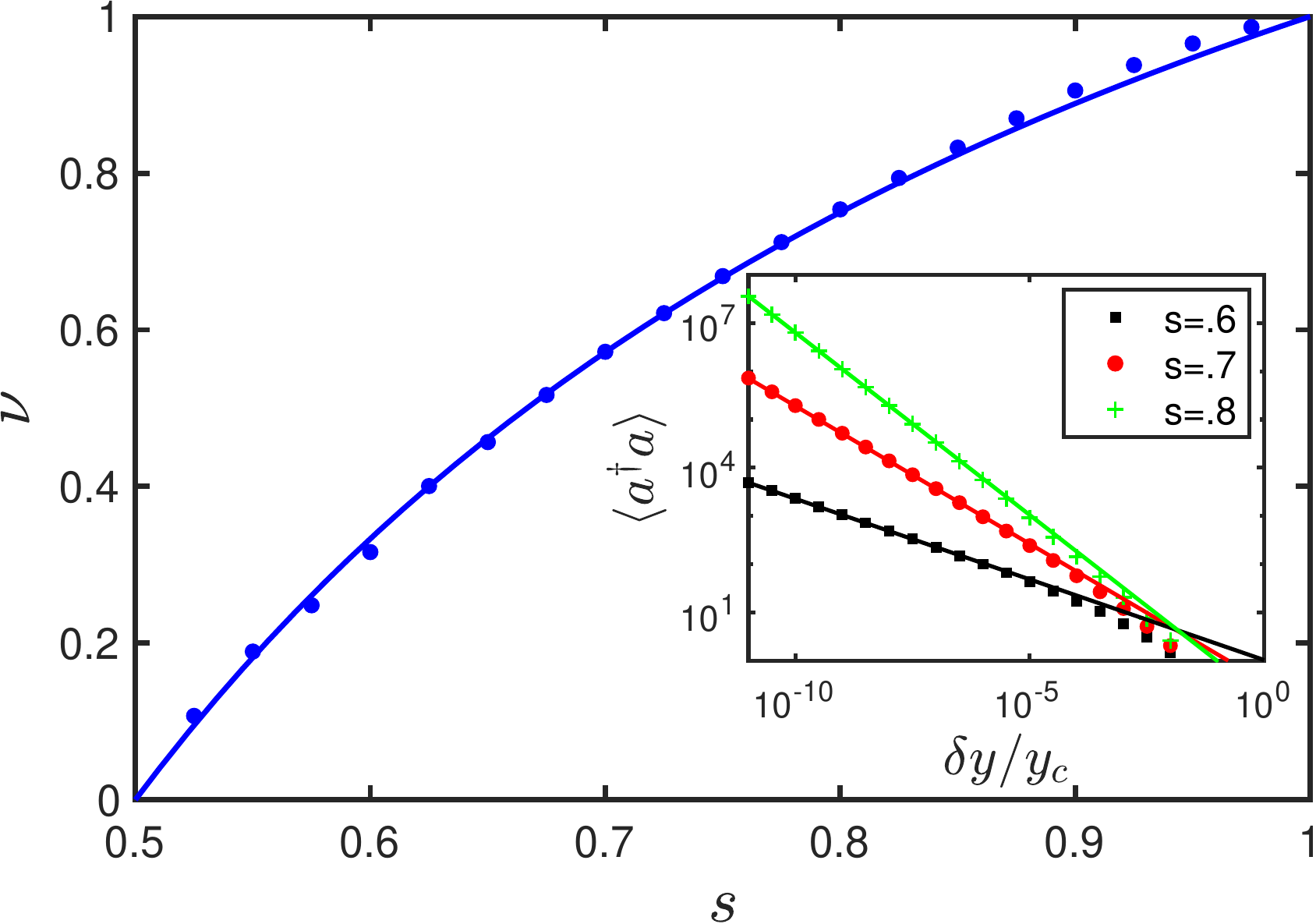}
  \caption{(color online) Photon-flux exponent versus the exponent of the colored bath. The (blue) dots are from the numerical integration of Eq.~\eqref{photoncorr} and the (blue) curve is our analytical prediction, $2-1/s$. The photon number does not diverge for $s<1/2$. Inset: Photon number as a function of distance from criticality for various $s$. The dots are data obtained from numerical integration of Eq.~\eqref{photoncorr} and the lines are linear fits from which the critical exponent is extracted. Here, the (black) squares and line are for $s=.6$, the (red) dots and line are for $s=.7$ and the (green) crosses and line are for $s=.8$. In this figure, we have chosen the parameters $\Delta=2\omega_b$, $\kappa=.5\omega_b$, $\gamma=.1\omega_b$, $T_b=0$, and $\mu_b=0$.}
  \label{photon_number}
\end{figure}

We first calculate the photon number \cite{2016RPPh...79i6001S},
\begin{equation}
n\equiv\langle a^\dagger a\rangle=\frac{1}{2}\int_{\omega} \langle  |a_{cl}(\omega)|^2\rangle-\frac{1}{2},
\label{photoncorr}
\end{equation}
at zero temperature and zero chemical potential. Near the critical point and at small frequencies, the photonic correlation function diverges and the integral in the last equation is dominated by its behavior near $\omega=0$. We can then safely replace $\langle a^*_{cl}(\omega)a_{cl}(\omega)\rangle$ by its low-frequency expression, $\langle x^*_{cl}(\omega)x_{cl}(\omega)\rangle$ (ignoring multiplicative factors, as we are interested in scaling relations), to analytically extract the critical exponent. We find (defining $\tilde{v}^2=v_I^2+v_R^2$)
\begin{align}
 n &\sim \int_{\omega} \left\langle  |x_{cl}(\omega)|^2\right\rangle
 \sim 
 \int_\omega  \frac{\kappa_{\rm eff}}{r^2+\tilde{v}^2\big|\frac{\omega}{\omega_z}\big|^{2s}-2r v_R\big|\frac{\omega}{\omega_z}\big|^s}\nonumber \\
&\propto r^{-2+1/s}.
\label{Eq. ph. flux exponent}
\end{align}
Using the fact that $r\propto \delta y$, one could also write $n\propto\delta y^{-(2-\frac{1}{s})}$. To derive Eq.~\eqref{Eq. ph. flux exponent}, we have used the low-frequency expression for the Keldysh Green's function, $-P^K_{x}(\omega)/(P^R_{x}(\omega,y)P^A_{x}(\omega,y))$; cf. Eqs.~\eqref{grcase3} and \eqref{DKcase1}. This yields the photon-flux exponent 
\begin{equation}
    \nu=2-\frac{1}{s},\qquad \mbox{for}\quad s>1/2.
    \label{correlationEXPmaincase}
\end{equation} 
We note that for $s<1/2$ the photon number does not diverge at the critical point. The fact that fluctuations do not diverge for sufficiently small $s$ is in harmony with the expectation that fluctuations are less important as the coupling becomes more long-ranged, in this case, along the temporal direction. The photon-flux exponent in the closely related models that we have considered in this work are presented in Table~\ref{CritEXPTABLE}. In Fig.~\ref{photon_number}, we compare this exponent against the numerical integration of Eq.~\eqref{photoncorr} and find excellent agreement. Furthermore, these values are consistent with the exponents obtained in Refs. \cite{PhysRevLett.115.043601,PhysRevA.94.063862}, as expected.

\subsection{Correlation function}
\begin{figure*}
\subfloat[][At criticality.]{
 \includegraphics[width=.43\linewidth]{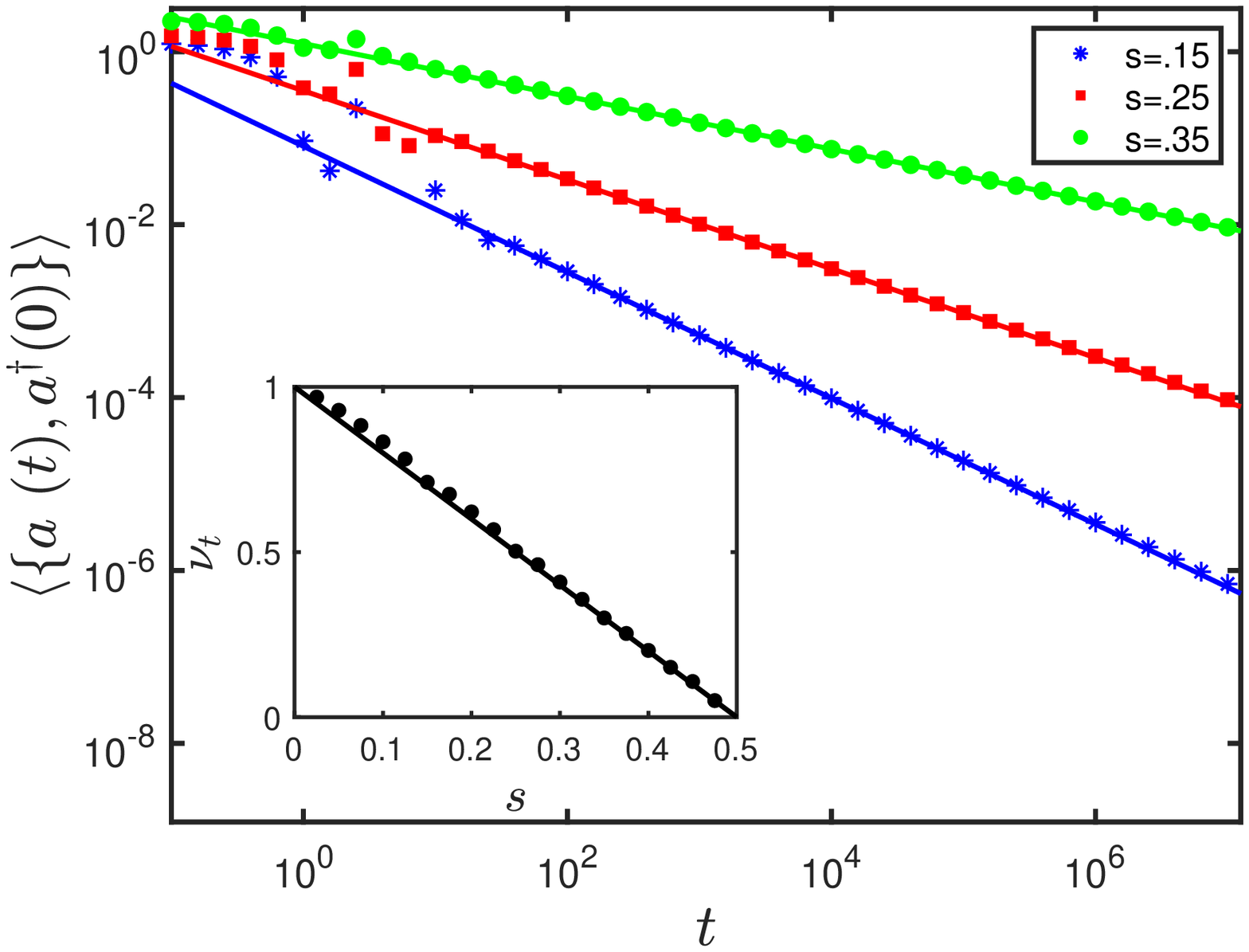}
\label{atcrit}
}\hspace{1cm}
\subfloat[][Away from criticality.]{
 \includegraphics[width=.43\linewidth]{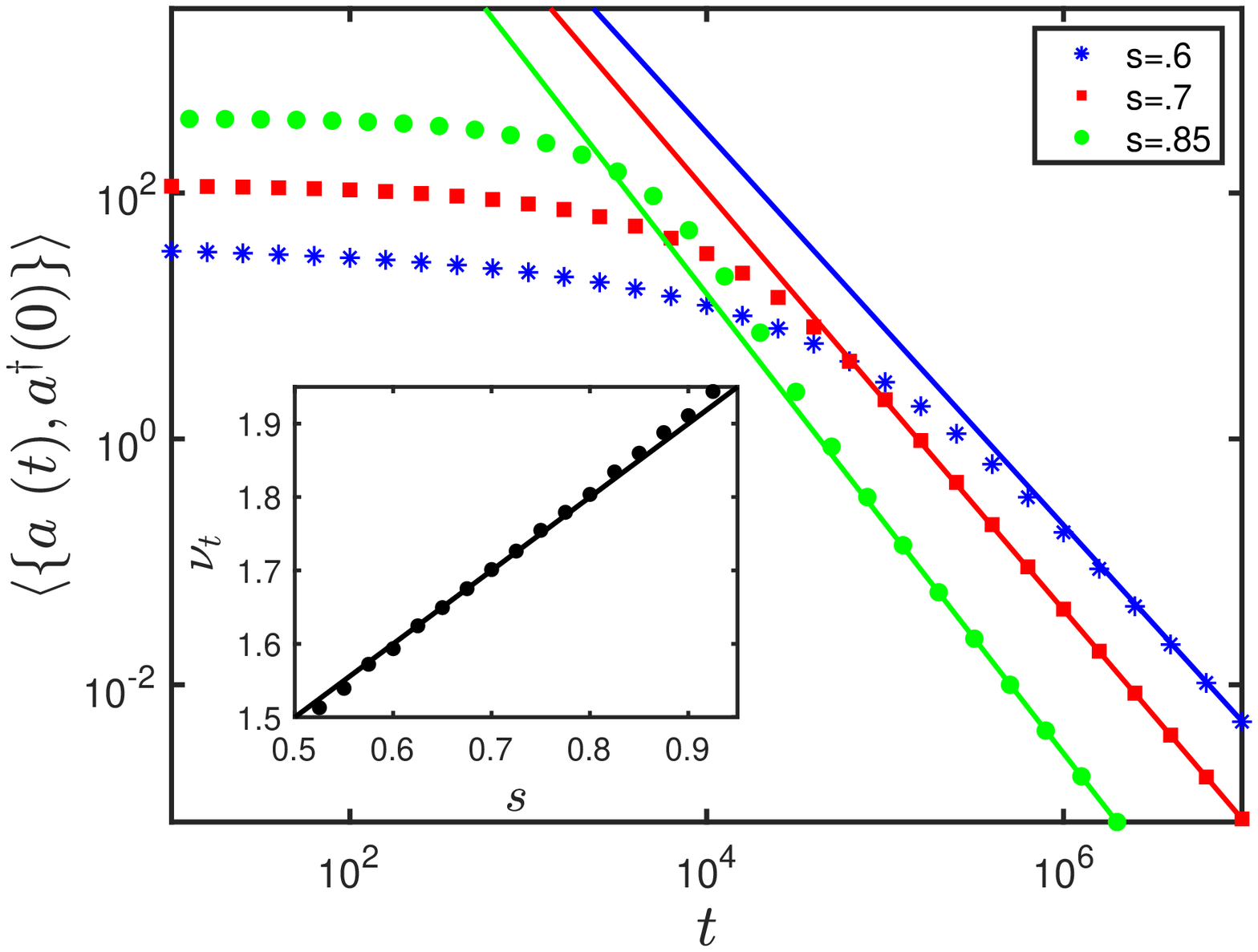}
\label{awayfromcrit}
}
\caption{(color online) Dynamical exponent. (a) Kelydsh Green's function as a function of time for various $s$ at criticality.  The colored markers are from numerical integration of the exact Kelydsh Green's function in frequency space and the solid lines are linear fits to late-time data. Inset: Dynamical exponent at criticality as a function of $s$. We find an excellent agreement between our analytical prediction of $1-2s$ and our numerical results. (b) Kelydsh Green's function as a function of time for various $s$ away from criticality (with $\delta y/y_c=10^{-4}$). Inset: Dynamical exponent away from criticality. We see excellent agreement between our analytical prediction (solid line) of $1+s$ and numerical results (markers).Here, we use the same parameters as in Fig.~\ref{photon_number}.}
\end{figure*}

We now turn to the auto-correlation function involving two different times. More specifically, we consider the Keldysh Green's function given by \cite{2016RPPh...79i6001S}
\begin{align}
iG^K_{\mathrm{ph}}(t-t')&\equiv\langle \{a(t),a^{\dagger}(t')\}\rangle=\langle a_{cl}(t)a_{cl}^*(t')\rangle \nonumber \\
&=\int_\omega\langle |a_{cl}(\omega)|^2\rangle e^{-i\omega (t-t')}.
\end{align}
Due to time translation symmetry, we can set $t'=0$. Near the critical point and at long times, we can again express the bosonic fields in terms of the critical field $x$ to obtain [using Eqs.~\eqref{grcase3} and \eqref{DKcase1}]
\begin{align}
G^K_{\mathrm{ph}}(t)\sim \int_\omega\frac{\kappa_{\rm eff}e^{-i\omega t}}{r^2+\tilde{v}^2\big|\frac{\omega}{\omega_z}\big|^{2s}-2rv_R\big|\frac{\omega}{\omega_z}\big|^s}.
\label{corrfunction}
\end{align}
There are two limits to consider, at or away from criticality, each of which yields a different dynamical critical exponent. Furthermore, away from criticality, there is a crossover from critical behavior at short times (though long compared to microscopic time scales) to non-critical behavior after a certain time which itself scales algebraically with the distance from the critical point; see Fig.~\ref{crossover}. This behavior is discussed in detail in Sec.~\ref{Critaway}.

\subsubsection{At criticality}
The correlation function at the critical point, $r=0$, at long times is given by
\begin{equation}
G^K_\mathrm{ph}(t)\propto\int_\omega\frac{e^{-i\omega t}}{\big|{\omega}\big|^{2s}}\propto \frac{1}{|t|^{1-2s}}.
\label{Eq. GK critical}
\end{equation}
This yields the dynamical exponent 
\begin{equation}\label{Eq. dyn exponent}
\nu_t=1-2s, \qquad \mbox{for} \quad s<1/2.    
\end{equation}
Note that this exponent only makes sense for $s<1/2$, 
while, for $s>1/2$, the integral is divergent due to the steep frequency dependence near $\omega=0$. Such divergence is, however, regulated at finite $N$ (see Sec.~\ref{sec:FS}). The exponent in Eq.~\eqref{Eq. dyn exponent} identifies the critical behavior when $s<1/2$ in the same regime where the photon-flux exponent is trivial (i.e., zero); of course, this is no coincidence and the two behaviors are tied together. In Fig.~\ref{atcrit}, we compare the dynamical critical exponent for several choices of the exponent $s<1/2$ against the exact numerical integration and find excellent agreement.

The infrared (IR) divergence for $s>1/2$ can also be cured by instead considering the \textit{mass-squared displacement} defined as $(\Delta x)^2\equiv\langle(x(t)-x(0))^2\rangle$ \cite{Weiss08}; this quantity is simply given by $G^K_\mathrm{ph}(t)-G^K_\mathrm{ph}(0)$. We find that
\begin{equation}
    \langle (x(t)-x(0))^2 \rangle \sim t^{2s-1},\quad \mbox{for} \quad s>1/2.
\label{Eq. anamolous}
\end{equation}
Interestingly, this equation suggests that the photon number (or rather its first quadrature) can be considered as a \textit{subdiffusive} particle with $(\Delta x)^2\sim t^\delta$ for an exponent $\delta=2s-1<1$. This behavior is indeed consistent with our conjectured fractional Fokker-Planck equation \eqref{FPEQ} in the absence of the external potential (i.e., $r=0$).

\subsubsection{Away from criticality}\label{Critaway}
Next we consider the noncritical case when $r\neq 0$. In the long-time limit, Eq.~\eqref{corrfunction} is dominated by small frequencies and is given by\footnote{In a careful evaluation of the integral, the integration contour should be deformed to one around the branch cut; see, for example, Ref.~\cite{PhysRevB.93.125128}.}
\begin{align}
G^K_\mathrm{ph}(t)
\propto \frac{1}{r^3|t|^{1+s}},
\label{Kaaway}
\end{align}
hence, the dynamical exponent $1+s$. Interestingly, the dynamical correlation function decays as a power law even away from criticality. Furthermore, the power-law decay is governed by the same exponent that controls the long-range temporal correlations in the bath. Therefore, one should think of this power-law as one that is directly inherited from the underlying long-range correlations in the bath, while the power-law dependence in Eq.~\eqref{Eq. GK critical} is genuinely due to the critical behavior at the critical point.
Same behavior is seen in long-range interacting models \cite{PhysRevB.93.125128}. In Fig.~\ref{awayfromcrit}, we compute the above exponent numerically and once again find excellent agreement between our analytical prediction and exact numerical integration.

\begin{figure}
  \centering\includegraphics[width=.45\textwidth, height=6cm]{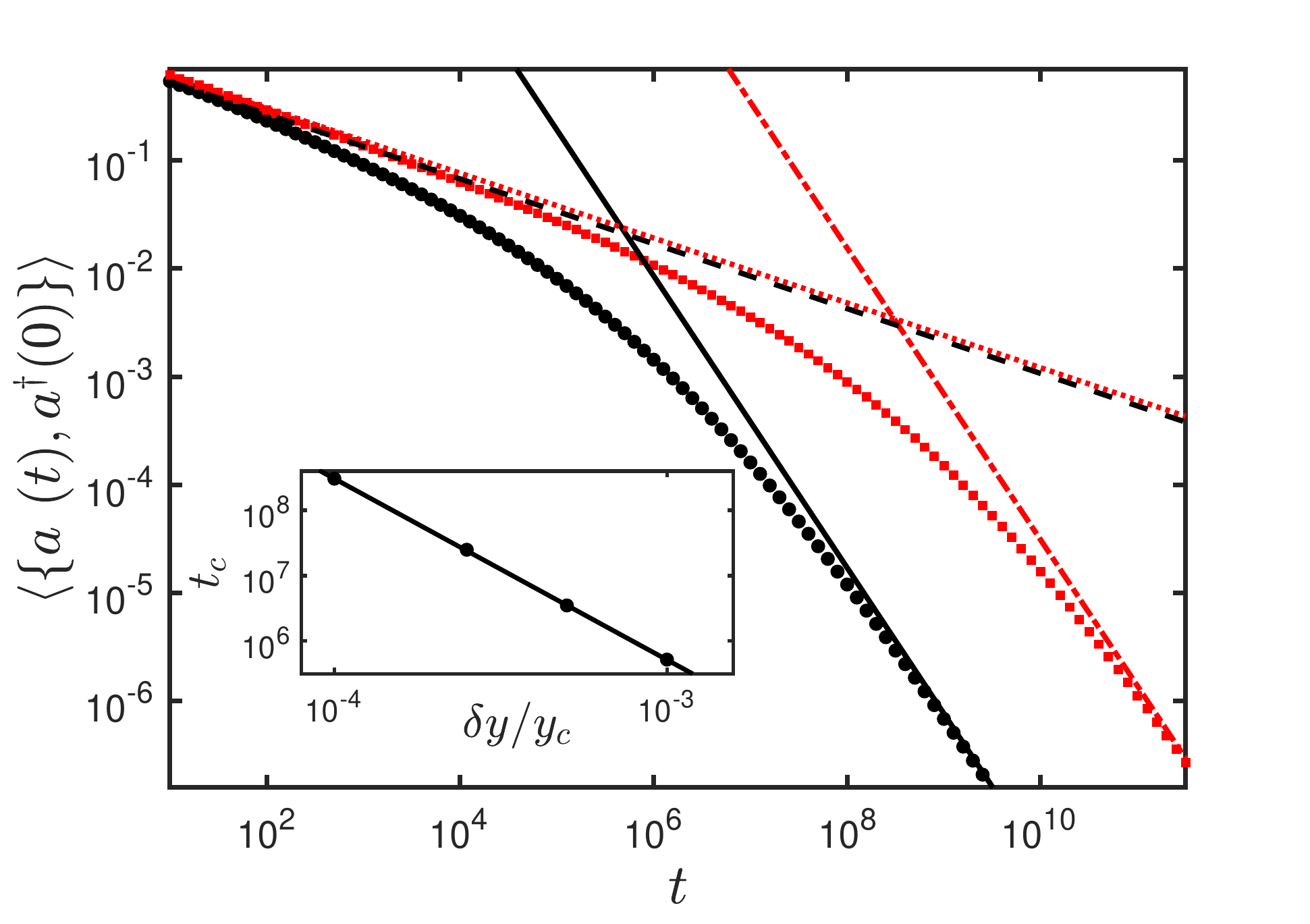}
  \caption{(color online) Crossover behavior of the dynamical correlation function for $s=.35$. The (black) dots are for $\delta y/y_c=10^{-3}$ and the (red) squares are for $\delta y/y_c=10^{-4}$. The (almost overlapping) dashed (black) line and dotted (red) line describe power-law decay with the dynamical exponent $\nu_t=1-2s$. The solid (black) line and the dotted-dashed (red) line describe power-law decay with dynamical exponent $\nu_t=1+s$. Inset: Crossover time, $t_c$, versus $\delta y$. $t_c$ is defined as the time when the solid and dashed lines intersect. The dots are from the numerical data and the line is a linear fit of this data. The crossover time exponent extracted from the slope of this line (approximately $2.77$) is in good agreement with the theoretical prediction of $1/.35\approx 2.86$. Again, we have used the same parameters as in Fig.~\ref{photon_number}.}
  \label{crossover}
\end{figure}

Even away from criticality, the auto-correlation function, $G^K_\mathrm{ph}(t)$ at short times decays with the same exponent that governs the critical behavior [cf. Eq.~\eqref{Eq. GK critical}]. In fact, one should expect a crossover from criticality at short times to noncritical behavior at long times. We have illustrated this behavior for the exponent $s=.35$ in  Fig.~\ref{crossover}. The crossover timescale can be estimated by finding the characteristic time where the critical and noncritical correlation functions in Eqs.~\eqref{Eq. GK critical} and \eqref{Kaaway} become comparable in magnitude (for an alternative derivation based on general scaling relations, see the end of Sec.~\ref{sec:FS}).  This occurs at a crossover time scale 
\begin{equation}
    t_c\sim r^{-1/s} \sim \delta y^{-1/s}.
\label{Eq. crossover time}
\end{equation}
In the inset of Fig.~\ref{crossover}, we plot the crossover time as a function of $\delta y$ and find excellent agreement between the theoretically predicted behavior and the exact numerical results. The crossover-time exponents for all other cases are presented in Table~\ref{crossovertable}.

\subsection{Response Function}
The response function is given by (for concreteness, we restrict ourselves to the retarded Green's function)
\begin{align}
iG^R_\mathrm{ph}(t)&\equiv \Theta(t)\langle [a(t),a^{\dagger}(0)]\rangle=
\langle a_{cl}(t)a_q^*(0)\rangle \nonumber \\
&=\int_\omega\langle a^*_{q}(\omega)a_{cl}(\omega)\rangle e^{-i\omega t}.
\end{align}
Again, near the critical point and at long times, the correlation function can be computed from the low-frequency expression in terms of the critical field $x$; we then find []using Eqs.~\eqref{grcase3} and \eqref{DKcase1}]
\begin{align}
G^R_\mathrm{ph}(t)\sim \int_\omega\frac{e^{-i\omega t}}{-r+\big|\frac{\omega}{\omega_z}\big|^{s}(iv_I \sgn(\omega)+v_R)}.
\label{rescorrfunction}
\end{align}
Again, there are two limits to consider, at or away from criticality, each yielding a distinct dynamical exponent; we designate them the response exponents to avoid any confusion with those describing the correlation function detailed in the previous section. 

We first consider the critical point where $r=0$, in which case we find
\begin{align}
G^R_\mathrm{ph}(t)
\propto \frac{\Theta(t)}{|t|^{1-s}},
\label{Eq. GR critical}
\end{align}
hence, the response exponent 
\begin{equation}
\nu'_t=1-s    
\end{equation}
On the other hand, away from criticality, $\delta y\neq 0$, we find
\begin{align}
G^R_\mathrm{ph}(t)\propto \frac{\Theta(t)}{r^2|t|^{1+s}}.
\label{Eq. GR noncritical}
\end{align}
Again, we see that, even away from criticality, the characteristic two-point function (response function, in this case) is governed by the same critical exponent as that of the long-range (temporal) correlations in the bath. And again, a comparison between Eqs.~\eqref{Eq. GR critical} and \eqref{Eq. GR noncritical} reveals a crossover from criticality to noncritical behavior at a time scale $t_c\sim \delta y^{-1/s}$. 

We note that for systems in (either global or effective) thermal equilibrium, there is a relationship between the response and dynamical exponents; see, for example, Ref.~\cite{tauber2014critical}. At finite temperature, this follows from the fluctuation-dissipation relation $G^R(t)=-\frac{1}{T}\Theta(t)\partial_tG^K(t)$, which dictates $\nu'_t=\nu_t-1$. On the other hand, at zero temperature, the response and dynamical exponents should be identical. In the present case, regardless of $\delta y$, we see that this relationship is not obeyed, a further indication that the system is not in thermal equilibrium either at or away from the critical point.

\subsection{Finite Temperature}\label{FiniteTempEXP}
\begin{figure}[t]
 \centering\includegraphics[width=0.45\textwidth, height=6.5cm]{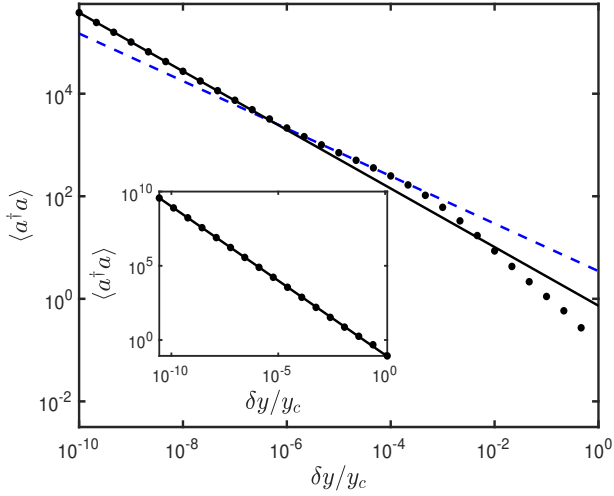}
 \caption{(color online) Effect of finite temperature. (a) Photon number versus distance from criticality for $s=.7$,~$T_b=\omega_b$ and $\mu_b=-.001\omega_b$. Other parameters are the same as the ones used in Fig.~\ref{photon_number}. The dashed-blue (solid-black) line is obtained from fitting the numerical data, represented by the black dots, between $\delta y/y_c=10 ^{-4} (10^{-8})$ and $\delta y/y_c =10 ^{-6} (10^{-10})$. Importantly, the correct critical exponent is extracted only sufficiently close to the critical point. Inset: $\langle a^\dagger a\rangle$ for $T_b=\omega_b$ and zero chemical potential, $\mu_b=0$. The critical exponent obtained from fitting the numerical data is approximately 1.0001, consistent with our analytical prediction, $\nu=1$.}
  \label{temp}
\end{figure}
Heretofore, we have only considered critical exponents for $T_b=0$ and $\mu_b=0$. We now consider the effect of finite temperature. (We remind the reader that, at $T_b=0$, the critical exponents are independent of $\mu_b$; see the discussion in Sec.~\ref{finiteTEMPTHEORY}.) Nagy and Domokos found that the photon-flux exponent decreases as a function of chemical potential at finite temperature (of the non-Markovian bath). In contrast, our low-frequency theory predicts that at finite temperature and chemical potential the photon-flux exponent should not change at all; cf. Sec.~\ref{MAINSECND}. We now resolve this apparent discrepancy. In Fig.~\ref{temp}, we plot the photon number versus the distance from criticality at finite chemical potential and temperature for $s=.7$. As a first fit, consider the dashed (blue)  line  obtained by fitting numerical data between $1- y/y_c =10 ^{-4}$ and $1-y/y_c =10 ^{-6}$ similar to Ref.~\cite{PhysRevLett.115.043601}; we then find an exponent ($\nu\approx.501$) consistent with the numerical result of Ref.~\cite{PhysRevA.94.063862}. Upon decreasing to small values of $1-y/y_c \sim 10 ^{-10}$, we see the critical exponent seems to increase. Indeed, the value of the critical exponent obtained from fitting data closer to criticality is in excellent agreement with our analytical prediction of $2-1/s$ (our numerically calculated exponent is approximately $.568$, while our analytically predicted exponent is approximately $.567$).  Thus, we are led to believe that the reason for the disagreement with Ref.~\cite{PhysRevA.94.063862} is that the authors of the latter reference did not consider close enough distances to criticality. Finally, we have numerically calculated the photon-flux exponent at zero chemical potential and finite temperature to verify our analytical prediction of $\nu=1$ [this can be analytically computed from Eq.~\eqref{c_aaction_case1}] regardless of the value of $s$; see inset of Fig.~\ref{temp}. Again, we find excellent agreement with the exact numerical calculation ($\nu\approx 1.0001$).

Finally, at finite temperature but zero chemical potential ($T_b\ne0$ and $\mu_b=0$), we analytically find [using Eq.~\eqref{c_aaction_case1}] that the dynamic and response exponents obey the fluctuation-dissipation relation, $\nu'_t=s=\nu_t-1$, indicating the system is indeed in thermal equilibrium as expected. This underscores the fact that the behavior of the system at $T_b\neq0$ and $\mu_b=0$ is fundamentally different from that at $T_b\neq0$ and $\mu_b\neq0$.

\subsection{Finite-Size Scaling: Statics and Dynamics}\label{sec:FS}
We now turn to finite-size effects and specifically aim to determine the finite-size scaling exponent, that is, how the photon number scales with the system size (spin \textit{length}) at the critical point. We explicitly derive the finite-size scaling exponent using a general scaling analysis~\cite{PhysRevA.87.023831}. In this section, we present explicit calculations in the presence of both Markovian and non-Markovian baths with $\mu_b,T_b=0$ (or when both $\mu_b,T_b\neq0$). The finite-size scaling exponent in the other cases considered in this work are presented in Table.~\ref{CritEXPTABLE}. Since the low-frequency theories describing both atomic and photonic fields are the same, we shall expect identical finite-size scaling exponents.

To this end, let us examine the interaction term, Eq.~\eqref{EFFECTIVEACTIONA}, under the scaling in Eq.~\eqref{case3scaling}. Upon this transformation, the strength of the interaction is rescaled as
\begin{equation}
    \frac{g_{\rm ph}}{N}\to \lambda^{3s-1}\frac{g_{\rm ph}}{N}.
    \label{Eq. g scaling}
\end{equation}
This scaling behavior immediately indicates that the ``lower critical exponent'' (using the analogy with the upper critical dimension) is given by $s_l=1/3$ below which the interaction term is irrelevant, and the Gaussian fixed point governs the critical behavior. In other words, for $s<1/3$, finite-size corrections can be ignored. For $s>1/3$, however, the finite-size scaling is nontrivial. As pointed out in Ref.~\cite{PhysRevA.87.023831}, this type of interaction term can be made invariant if $N$ scales with $\lambda$ in a certain way. More specifically, this term can be made invariant if $N$ itself is scaled as $N\to \lambda^{3s-1}N$. Let us recall that the photon number itself scales as $\lambda^{2s-1}$; this simply follows from the scaling of the classical field, $x_{cl} \to \lambda^{s-1/2}x_{cl}$, together with the fact that the critical scaling of the photon number is captured by the fluctuations of the critical field $x$, i.e., $\langle a^\dagger a\rangle\sim\langle x_{cl}^2\rangle$. Combining the two scaling laws, we find that, at the critical point, 
\begin{equation}
\langle a^\dagger a \rangle \propto N^{(2s-1)/(3s-1)},
\end{equation}
thus identifying the finite-size scaling of the photon number as
\begin{equation}
    \alpha=\frac{2s-1}{3s-1}, \quad \mbox{for}\quad s>1/2.
\label{Eq. finite-size}
\end{equation}
For $s=1$, this correctly reproduces the result of Ref.~\cite{PhysRevA.87.023831} as expected since the retarded Green's function becomes the standard one with the frequency dependence appearing linearly. The exponent reported in Eq.~\eqref{Eq. finite-size} suggests that, for $s<1/2$, the photon number does not diverge with the system size $N$, a fact that is also consistent with the results presented in Sec.~\ref{PFEXP} where we examined the dependence of photon number on the distance from the critical point. 

The situation described above poses a dilemma: On the one hand, we observed from the finite-size scaling of the interaction parameter that the lower critical exponent is $s_l=1/3$. On the other hand, there does not appear to be any finite-size scaling within the range $1/3<s<1/2$. It turns out that the solution of this puzzle lies in the dynamics. To this end, we first describe a general scaling analysis that not only unifies our treatment of various critical properties, but also sheds light on the finite-size scaling of the dynamics.

The scaling behavior of the relevant terms in the action [i.e., the sum of Eqs.~\eqref{caseIrealtimeaction} and \eqref{EFFECTIVEACTIONA}] reveals that the action is invariant under 
\begin{equation}
\begin{split}
t\,\,\rightarrow \,\,&\lambda\, t, \\
x_{cl}(t)\rightarrow \lambda^{s-1/2} x_{cl}(t),\qquad 
& x_{q}(t)\rightarrow\frac{1}{\sqrt{\lambda}}x_{q}(t), \\
r \rightarrow \lambda^{-s}r,\qquad &N\rightarrow\lambda^{3s-1} N,
\end{split}
\label{finiteNscaling}
\end{equation}
where we have put together the scaling transformation in Eq.~\eqref{case3scaling} together with the system-size scaling [cf. Eq.~\eqref{Eq. g scaling}] and an appropriate scaling for the mass term. 
To be concrete, let us consider the auto-correlation function $\langle x_{cl}(t)x_{cl}(0)\rangle$ at or near the critical point at a finite system size $N$; one can also consider the response function without changing the main conclusions. Using the above scaling behavior, the correlation function should satisfy the scaling relation
\begin{align}
\langle x_{cl}(t)x_{cl}(0)\rangle=\lambda ^{1-2s}C\!\left(\lambda |t|,\lambda^{-s}r,\frac{1}{\lambda^{3s-1}N}\right),
\label{lambdadynamicalscalling}
\end{align}
where $C$ is a general scaling function. This scaling behavior already reveals a few properties that we have encountered before. First, at the critical point ($r=0$) and in the thermodynamic limit ($N\to\infty$), we can set $\lambda=t^{-1}$ to obtain 
\begin{equation}
\langle x_{cl}(t)x_{cl}(0)\rangle= \frac{C(1,0,0)}{|t|^{1-2s}}
\propto \frac{1}{|t|^{1-2s}},
\end{equation}
consistent with Eq.~\eqref{Eq. GK critical}. Of course, this makes sense for $s<1/2$, while, for $s>1/2$, we can instead use the mean square displacement to arrive at a similar scaling consistent with the anomalous diffusion in Eq.~\eqref{Eq. anamolous}. Also, we can  easily identify the photon-flux exponent by setting $t=0$ in the thermodynamic limit ($N\to\infty$) while choosing $\lambda=r^{1/s}$ to find
\begin{equation}
\left\langle x_{cl}^2\right\rangle= \frac{C(0,1,0)}{r^{2-1/s}} 
\propto \frac{1}{r^{2-1/s}}.
\end{equation}
Again, this is consistent with Eq.~\eqref{Eq. ph. flux exponent} for $s>1/2$.
We can even determine the crossover time scale and its scaling behavior in the thermodynamic limit ($N= \infty$). In this limit, we find (with $\lambda=t^{-1}$)
\begin{equation}
    \langle x_{cl}(t)x_{cl}(0)\rangle = \frac{C(1,|t|^{s}r,0)}{|t|^{1-2s}} \equiv
    \frac{C_c(|t|^{s}r)}{|t|^{1-2s}},
\end{equation}
where, in the last line, we have defined the scaling function $C_c$. A reasonable assumption is that this function exhibits a crossover when its argument is of the order of 1; smaller values of the argument describe the short-time (critical) behavior, while larger values represent long-time (non-critical) behavior. 
The latter condition ($|t|^{s}r\sim 1$) then determines the crossover time scale as $t_c\sim \delta y^{-1/s}$, which is again consistent with Eq.~\eqref{Eq. crossover time}. 

\begin{table}
\begin{tabular}{|c|c|c|}\hline
& Finite $\delta y$ & Finite $N$   \tabularnewline
\hline
\makecell{Characteristic \\time scales} & $t\sim\delta y ^{-\zeta_c}$ & $t\sim N ^{\zeta}$ \tabularnewline
\hline 
\makecell{MB on $\&$ NMB on \\ $T_b= 0,~\mu_b\leq 0$}  & $1/s$ & $\begin{cases}
    1/(3s-1),\quad &s>1/3\\
    0,&s<1/3 
  \end{cases}$  \tabularnewline
\hline
\makecell{MB on $\&$ NMB on \\ $T_b\neq 0,~\mu_b=0$}  &  $1/s$ & $1/(2s)$   \tabularnewline
\hline
\makecell{MB on $\&$ NMB off \\ $T_b= 0,~\mu_b\leq 0$}  & 1 & $1/2$   \tabularnewline
\hline
\makecell{MB off $\&$ NMB on \\ $T_b= 0,~\mu_b\leq 0$}  &  $1/s$  &$\begin{cases}
    1/(2s-1),&s>1/2\\
    0,&s<1/2
  \end{cases}$   \tabularnewline
\hline
\end{tabular}
\caption{Table of the characteristic time scales for the models considered in this work. The exponent $\zeta_c$ corresponds to the cross-over time away from criticality ($\delta y\ne 0$ in the limit $N\to\infty$), while the exponent $\zeta$ characterizes the dynamics' finite-size scaling at criticality ($\delta y =0$ at finite $N$).}
\label{crossovertable}
\end{table}

The scaling relation \eqref{lambdadynamicalscalling} can also be used to determine the finite-size effects. Setting $t=0$ and $r=0$ and choosing $\lambda=N^{-1/(3s-1)}$, we find 
\begin{equation}
\left\langle x_{cl}^2\right\rangle= {C(0,0,1)} N^{(2s-1)/(3s-1)},
\end{equation}
where the scaling behavior is consistent with Eq.~\eqref{Eq. finite-size}. This result displays the finite-size scaling in the static (equal-time) distribution of the photonic field. 

Importantly, finite-size effects leave distinct fingerprints in the dynamics as well. 
To this end, let us keep $t$ finite, but set $\delta y=0$, and choose $\lambda=t^{-1}$. The scaling relation~\eqref{lambdadynamicalscalling} then yields 
\begin{align}
\langle x_{cl}(t)x_{cl}(0)\rangle
&= \frac{1}{|t|^{1-2s}}
C\left(1,0,\frac{t^{3s-1}}{N}\right) \nonumber \\
&\equiv \frac{1}{|t|^{1-2s}} 
\tilde{C}\left(\frac{t}{N^{1/(3s-1)}}\right),
\end{align}
where, in the last line, we have defined the scaling function $\tilde C$.
This then leads to a new time scale $t_N \sim N^{\zeta}$ with $\zeta=1/(3s-1)$ characterizing the finite-size effects. 
Notice that, at the critical point ($\delta y=0$) in the thermodynamic limit ($N \to \infty$), the correlations are scale invariant ($\sim 1/|t|^{1-2s}$) and there is no characteristic time scale. In a finite system, however, a characteristic time scale should emerge that is sensitive to, but diverges with, the system size. While we would not further explore the precise nature of these finite-size effects, the system is likely to exhibit damped oscillations over a time scale set by $t_N$.

Finally, we are in a position to explain the dilemma posed earlier: While static (i.e., equal-time) correlations, and specifically the photon-flux exponent, are insensitive to the finite system size for $s<1/2$, the characteristic time scale $t_N$ exhibits a nontrivial finite-size scaling in the entire range of $1/3<s<1$. 
Remarkably, this implies that, even when interactions (i.e., finite-size effects) may be entirely ignored as far as static properties are concerned, they can nontrivially affect the dynamics; in our model, this occurs when $1/3<s<1/2$. On the other hand, when $s<1/3$, the critical behavior of the model----both statics and dynamics---is fully described by a Gaussian fixed point.

We present the exponents characterizing the crossover time (off criticality) as well as  the finite-size time scale (at criticality) in this and the closely related models in Table~\ref{crossovertable}.

\section{Closely Related Models}\label{sec:othermodels}
In this section, we discuss other closely related models to place our results in some context. We stress that our non-trivial basis rotation (see Fig.~\ref{rotation_fig}) is also helpful in understanding these cases. More specifically, we consider two cases where the Markovian bath is present/absent while the non-Markovian bath is absent/present, respectively. We show that these two cases yield different low-frequency field theories and critical properties. For each case, we calculate the retarded and Keldysh Green's functions of the low-frequency theory, the distribution function, and discuss the nature of their phase transition. These results are summarized in Table~\ref{situtionTable}, along with the previous two cases we have discussed. The critical exponents of all limits we consider are presented in Tables~\ref{CritEXPTABLE} and \ref{crossovertable}.

\subsection{Markovian bath on $\&$ non-Markovian bath off}\label{CASEMBONNMBOFF}
We first consider the case when the non-Markovian bath is absent ($\gamma=0$) and the system is at zero temperature to demonstrate the usefulness of our approach. This scenario describes the standard setting of the open Dicke model, and has been considered in numerous other works; see, for example, Refs.~\cite{PhysRevA.84.043637,PhysRevA.87.023831,2012NJPh...14h5011O,2018arXiv180509828K}. In the low-frequency limit, the inverse retarded and Keldysh Green's functions of the $x$ field are
\begin{align}
P^R_{x}(\omega)\approx -r +i\omega \chi,~~~P^K_{x}(\omega)=2i\kappa_{\rm eff}.
\label{GREENSFUNCTIONMBONNMBOFF}
\end{align}
Note that this is the same inverse Keldysh Green's function as in the last section. The distribution function is given by $F_{x}(\omega)=\frac{\kappa(1+\chi^2)}{2\omega \chi }$, indicating that the photonic mode is effectively in equilibrium at a temperature of $T_{\rm eff}=\frac{\kappa^2+\Delta^2}{4\Delta}$; this is in agreement with the effective temperature reported in Ref.~\cite{PhysRevA.87.023831}. We stress that the effective thermal equilibrium governs the low-frequency properties of the system \cite{PhysRevA.87.023831}. 
Furthermore, we find the same effective temperature for the atomic field; see App.~\ref{sec:LEATOMICFIELDcaseIII}.
This further indicates that the entire system of photons coupled to atoms is effectively in global thermal equilibrium.

The nature of this phase transition was investigated in detail by Dalla Torre \etal \cite{PhysRevA.87.023831} where it was shown that the phase transition is classical and in the same universality class as the classical Ising model with infinite-range interactions (and no conserved quantities), a mean-field version of model A of Halperin-Hohenberg classification \cite{Hohenberg77}. Considering a different type of atomic dissipation, however, Ref.~\cite{PhysRevA.87.023831} reported different effective temperatures for atoms and photons. For completeness, we explicitly present the scaling argument in the above reference. Rewriting the action at the critical point in the time domain, we have
\begin{align}
S_{x}^{\rm eff}= \int_t 
(x_{cl},x_{q})
\bmx
0 &  \chi\partial_t \\
-\chi\partial_t & 2i\kappa_{\rm eff}
\emx
\left(\begin{array}{cc}
x_{cl} \\
x_{q}
\end{array}\right).
\label{c_aaction_case1}
\end{align}
One can easily see that this equation is invariant under the rescaling 
\begin{equation}
\begin{split}
t\,\,\rightarrow\,\,& \lambda\, t, \\
x_{cl}(t)\rightarrow \sqrt{\lambda} x_{cl}(t), \qquad &
x_{q}(t)\rightarrow\frac{1}{\sqrt{\lambda}}x_{q}(t).
\end{split}
\end{equation}
Again the scaling dimension of the quantum field is more negative than that of the classical field, rendering ``quantum interaction vertices'' irrelevant and the  phase transition classical in nature \cite{PhysRevA.87.023831}. The critical exponents in this case are presented in Table~\ref{CritEXPTABLE}. In particular, we obtain a photon-flux exponent of $\nu=1$ and a finite-size scaling exponent of $\alpha=1/2$~\cite{PhysRevA.84.043637,PhysRevA.87.023831,2018arXiv180509828K}. Note that we found the same exponents in the case considered in Sec.~\ref{FiniteTempEXP}. In fact, this is simply because the phase transition in both limits is classical and thermal. Here too, we find $G_{\mathrm{ph}}^R(t)=-\frac{1}{T_{\rm eff}}\Theta(t)\partial_tG_\mathrm{ph}^K(t)$, indicating that the system is in effective thermal equilibrium.

\subsection{Markovian bath off $\&$ non-Markovian bath on}\label{sec:othermodelscaseIII}
We now consider the limit where the Markovian bath is absent ($\kappa=0$) while the non-Markovian bath is present; we further assume that the latter bath is at zero temperature ($T_b=0$) while the  chemical potential is arbitrary ($\mu_b\leq 0$). In this case, the critical point is given by $y_c=\sqrt{\Delta\omega_z}$. Furthermore, in the limit $\kappa=0$, we have $\theta=\pi$, thus the $z$ field (which is, by definition, always gapped) is just the imaginary part of the $a$ field (see Fig.~\ref{rotation_fig}). We note that these features hold even when both the temperature and the chemical potential are finite (see the discussion at the end of Sec.~\ref{MAINSECND}). Green's function of the $x$ field is given by 
\begin{align}
P^R_{x}(\omega)\approx -r +
\left|\frac{\omega}{\omega_z}\right|^s\bigg(iv_I\sgn(\omega)-v_R\bigg),
\end{align}
while the inverse Keldysh Green's function becomes
\begin{align}
P^K_{x}(\omega)\approx 2iv_I\left|\frac{\omega}{\omega_z}\right|^s .
\end{align}
The distribution function is then given by $F_{x}(\omega)=\sgn(\omega)$, signalling that the photonic mode is in equilibrium at zero temperature. In fact, it is easy to see that the fluctuation-dissipation relation holds at all frequencies, indicating that the system is genuinely in equilibrium. This might sound surprising given the time dependence of the Hamiltonian \eqref{Eq. H}; however, this time dependence can be gauged away without affecting the dynamics. In contrast, in the presence of the Markovian bath, the system is genuinely out of equilibrium even in the rotating frame since the dynamics in the presence of both the (Markovian) dissipation and the (time-independent) Hamiltonian violates the fluctuation-dissipation relation. 

Next we investigate the nature of the phase transition in this model. Rewriting the action in the time domain, we have 
 \begin{align}
S_{x}^{\rm eff}=& \int_t x_q(t) (-v \partial_t^s+r )x_{cl}(t)\nonumber \\ &-ib_sv\int_t\int_{t'} \frac{x_{q}(t)x_{q}(t')}{|t-t'|^{1+s}},
\label{c_aaction_case2}
\end{align}
with $b_s=(2/\pi)\sin^2(\frac{\pi s}{2})\Gamma(1+s)$. One can easily see that, at the critical point ($r=0$), the action is scale-invariant upon the rescaling
\begin{equation}
    \begin{split}
t\,\,\rightarrow &\,\,\lambda\, t, \\
x_{cl}(t)\rightarrow \lambda^{\frac{s-1}{2}} x_{cl}(t), \qquad &
x_{q}(t)\rightarrow \lambda^{\frac{s-1}{2}}x_{q}(t).
    \end{split}
    \label{case2scaling}
\end{equation}
Thus, the quantum and classical fields have the same scaling dimension,
rendering the phase transition quantum in nature \cite{RevModPhys.59.1}. 
This is not surprising as the system is genuinely in equilibrium at zero temperature.
To gain further insight into the quantum nature of this model, let us examine the behavior of the effective interaction terms under rescaling. We find that a representative interaction term scales as 
\begin{equation}
S_{{\rm int},x}\sim\lambda^{2s-1}\frac{1}{N}\int_t x^4(t).
\label{interactionscaseIV}
\end{equation}
Here, we have dropped the quantum/classical labels as they have the same scaling dimensions. More precisely, the term with three quantum fields and one classical field that was previously neglected in Eq.~\eqref{EFFECTIVEACTIONA} is just as relevant as the term with three classical fields and one quantum field. Specifically, for $s>1/2$, the interaction term is relevant and determines the finite-size scaling at the critical point. As the action now contains a relevant term with more than two quantum fields, we cannot find an equivalent description in terms of a classical Langevin equation.

The above discussion indicates a lower critical exponent of $s_l=1/2$ below which interactions can be neglected. For $s>1/2$, the interaction term can be made scale invariant by rescaling $N$ by a factor of $\lambda^{2s-1}$. By similar arguments to those in Sec.~\ref{sec:FS}, one can see that the photon number does not diverge with $N$ for any $s<1$ even at the critical point. Yet again, following arguments similar to those in Sec.~\ref{sec:FS}, we find that the dynamics exhibits a nontrivial finite-size scaling for $s>1/2$ where a characteristic time scale emerges as $t_N \sim N^{1/(2s-1)}$. We thus see that the finite-size scaling affects statics and dynamics differently in both equilibrium and non-equilibrium settings.

The critical exponents of the model are presented in Table~\ref{CritEXPTABLE}. Notably, we observe that the dynamic and response exponents are the same, consistent with the fact that the system is in equilibrium at zero temperature \cite{tauber2014critical}. The critical exponents of this model are different from their counterparts in the other models that we have considered before. Most importantly, the phase transition here is quantum in nature while all the other models exhibit classical phase transitions. Furthermore, the model here is at zero temperature, while, in the other models we have considered, either an effective thermal equilibrium follows that is characterized by an effective temperature or a genuinely non-equilibrium (but classical) behavior emerges distinct from both zero and finite temperature.

\section{Summary and Conclusion}\label{sec:FUTURE}
In this work, we have investigated a variant of a driven-dissipative Dicke model where the atomic mode is coupled to a colored bath while the cavity mode is coupled to a Markovian bath, a model that was originally introduced by Nagy and Domokos and was interestingly found to have critical exponents that vary with the spectral density of the colored bath \cite{PhysRevLett.115.043601,PhysRevA.94.063862}. In this work, we have first derived the effective low-frequency Schwinger-Keldysh field theory of this model, which is made possible by a non-trivial basis rotation that allowed us to identify and integrate out the massive modes. Using this low-frequency field theory, we have analytically calculated various critical exponents and discussed the nature of the non-equilibrium phase transition. These results, along with those for other closely related models, are summarized in Tables~\ref{situtionTable},~\ref{CritEXPTABLE} and~\ref{crossovertable}. An important conclusion of our paper is that the corresponding non-equilibrium phase transition is classical in nature, not quantum as previously claimed. We have also compared the analytical expressions for the critical exponents against exact numerical calculations and found excellent agreement. 
Specifically, our analytical expression for the photon flux exponent is in excellent agreement with the numerical results obtained by Nagy and Domokos. We also resolved a discrepancy between our results and those in Ref.~\cite{PhysRevA.94.063862} which reported that critical exponents change as a function of finite chemical potential at finite temperature. In contrast, our investigation close enough to criticality has revealed that the critical exponents are independent of the chemical potential.

There are several interesting open questions for future investigation. 
It appears that the Beliaev process does not 
generate a coupling to a non-Markovian bath \cite{PhysRevA.98.063608} as it was originally proposed \cite{PhysRevLett.115.043601,PhysRevA.94.063862}. 
Therefore, an important future direction is to identify scenarios where
a non-Markovian bath emerges, or alternatively can be engineered, for the atomic modes. In this work, we have also conjectured a fractional Fokker-Planck equation for the probability distribution function of the photonic (or the atomic) field. It would be interesting to formally derive this equation from our effective theory. Another interesting though challenging direction is to perform a direct numerical simulation of the original model of photons coupled to atoms to access various critical properties including finite-size scaling exponents; however, this would be particularly  challenging as it requires solving a 
generalized many-body master equation in the form of a integro-differential equation \cite{RevModPhys.88.021002}. Finally, identifying non-equilibrium quantum critical behavior in driven-dissipative systems (currently missing from Table \ref{situtionTable}) is still a challenge even with the aid of the long-time memory in the bath \cite{PhysRevLett.122.110405,2010NatPh...6..806D,PhysRevB.85.184302}.

\begin{acknowledgments}
We thank A. Chakraborty, J. T. Young, J. Curtis, D. Nagy and P. Domokos for helpful discussions. R.L. and A.V.G. acknowledge support by DoE BES QIS program (award No. DE-SC0019449), DoE ASCR Quantum Testbed Pathfinder program (award No. DE-SC0019040), NSF PFCQC program, AFOSR, ARO MURI, ARL CDQI, and NSF PFC at JQI. M.F.M.  acknowledges support
from NSF under Grant No. DMR-1912799 and start-up funding from Michigan State University.
\end{acknowledgments}

\appendix
\section{low-frequency Theory of Atomic Field}\label{sec:LEATOMICFIELD}
In this section, we present the low-frequency theory of the atomic field. Integrating out the cavity degrees of freedom in Eq.~\eqref{FULLACT}, we find (see also \cite{PhysRevA.94.063862}),
\begin{equation}
S_{\mathrm{at}}^{\mathrm{eff}} = \int_\omega
\b{v}_{\mathrm{at}}^\dagger 
\bmx
 0 & \b{P}^A_{\mathrm{at}}(\omega) \\
\b{P}^R_{\mathrm{at}}(\omega)& \b{P}^K_{\mathrm{at}}(\omega)
\emx
 \b{v}^{\phantom{\dagger}}_{\mathrm{at}}\,,
 \label{eq:S_eff_b}
\end{equation}
where 
\begin{equation}
\b{v}_\mathrm{at}^\dagger(\omega)=(b_{cl}^*(\omega),b_{cl}(-\omega),b_{q}^*(\omega),b_{q}(-\omega)),
\end{equation}
and $\b{P}^R_{\rm at}(\omega)$, $\b{P}_{\rm at}^A(\omega)$, and $\b{P}^K_{\rm at}(\omega)$ are $2\times 2$ matrices. As done in the main text for the photonic fields, we have absorbed a factor of 1/2 into the atomic fields to match the notation of Refs.~\cite{PhysRevLett.115.043601,PhysRevA.94.063862}. This factor of 1/2 was missing in Refs.~\cite{PhysRevLett.115.043601,PhysRevA.94.063862}\label{factortwo} (see Ref.~\cite{2016RPPh...79i6001S}), but does not effect the critical properties of the atomic field. The inverse retarded and advanced Green's functions of the atomic field are given by
\begin{align}
&\b{P}^R_{\mathrm{at}}(\omega)=(\b{P}^A_{\rm at}(\omega)^{-1})^\dagger
\nonumber \\ &
=\bmx
 P^R_\mathrm{at}(\omega)+\Sigma_\mathrm{at}^R(\omega) &\Sigma_\mathrm{at}^R(\omega)   \\
\Sigma_\mathrm{at}^R(\omega)  & P^A_\mathrm{at}(-\omega)+\Sigma_\mathrm{at}^R(\omega)
\emx,
\end{align}
where the atomic self-energy, $\Sigma_\mathrm{at}^R(\omega)$, is given by
\begin{equation}
\Sigma_\mathrm{at}^R(\omega)=-\frac{y^2}{4}\bigg(\frac{1}{P^R_\mathrm{ph}(\omega)}+\frac{1}{P^A_\mathrm{ph}(-\omega)}\bigg).
\end{equation}
The inverse Keldysh Green's function, describing the effective bath for photons, is given by
\begin{align}
{\b{P}}^K_{\mathrm{at}}(\omega)=
\bmx
P^K_\mathrm{at}(\omega)+g(\omega) &  g(\omega)  \\
g(\omega)  & P^K_\mathrm{at}(-\omega)+g(\omega)
\emx,
\end{align}
where
\begin{equation}
g(\omega)=\frac{y^2}{4}\bigg(\frac{P^K_\mathrm{ph}(\omega)}{|P^R_\mathrm{ph}(\omega)|^2}+\frac{P^K_\mathrm{ph}(-\omega)}{|P^R_\mathrm{ph}(-\omega)|^2}\bigg).
\end{equation}

Similar to our treatment of the cavity mode in the main text, we change our basis to two real fields. In this case too, we seek a basis where $\b{P}^R_{\mathrm{at}}(\omega)$ becomes diagonal at zero frequency ($\omega=0$) and close to the critical point; at the critical point ($y=y_c$), one of the eigenvalues must vanish due to the emergence of a soft mode. To proceed, we write $\b{P}^R_{\mathrm{at}}(\omega,y)$ as 
\begin{align}
\b{P}^R_{\mathrm{at}}(\omega,y)&=\b{P}^R_{\mathrm{at}}(0,y_c)+ \\
&\bmx
\delta\Sigma_\mathrm{at}^R+\omega-K^R(\omega)  & \delta\Sigma_\mathrm{at}^R  \\
\delta\Sigma_\mathrm{at}^R  & \delta\Sigma_\mathrm{at}^R-\omega-K^A(-\omega) 
\emx, \nonumber
\end{align}
where $\delta\Sigma_\mathrm{at}^R=\Sigma_\mathrm{at}^R(\omega,y)-\Sigma_\mathrm{at}^R(0,y_c)$; here, we have explicitly included the dependence of $\b{P}^R_{\mathrm{at}}(\omega)$ and $\Sigma_\mathrm{at}^R(\omega)$ on $y$. Again, this expression is exact and no approximations have been made yet [given Eq.~\eqref{eq:S_eff_b}]. The set of real fields of classical and quantum fields that diagonalizes $\b{P}^R_{\mathrm{at}}(0,y_c)$ are given by
\begin{align}
\left(\begin{array}{cc}
\phi_{cl/q}(\omega) \\
 \zeta^*_{cl/q}(-\omega) \end{array}\right)=\b{R}_{\mathrm{at},cl/q}\left(\begin{array}{cc}b_{cl/q}(\omega) \\
 b^*_{cl/q}(-\omega) \end{array}\right),
\label{bfieldtransform1}
\end{align}
where
\begin{align}
\b{R}_{\mathrm{at},cl/q}=\left(\begin{array}{cc} 1 & 1  \\
 i & -i
\end{array}\right).
\end{align}
This corresponds to setting $\theta=\pi$ in Fig.~\ref{rotation_fig} (upon relabeling the axes). In other words, $\phi$ and $\zeta$ are simply the real and imaginary part of the $b$ field, respectively; a similar identification has been introduced in Refs.~\cite{Drummond05,Dechoum16}. The action in this new basis is then
\begin{equation}
S_\mathrm{at}^{\mathrm{eff}} = \int_\omega
\b{\tilde{v}}_{\mathrm{at}}^\dagger 
\bmx
 0 & \b{\tilde{P}}^A_{\mathrm{at}}(\omega) \\
\b{\tilde{P}}^R_{\mathrm{at}}(\omega) & \b{\tilde{P}}^K_{\mathrm{at}}(\omega)
\emx
 \b{\tilde{v}}^{\phantom{\dagger}}_{\mathrm{at}}\,,
 \label{eq:S_eff_b_newbasis}
\end{equation}
where 
\begin{equation}
\b{\tilde{v}}_\mathrm{at}^\dagger(\omega)=(\phi_{cl}^*(\omega),\zeta^*_{cl}(\omega),\phi_{q}^*(\omega),\zeta^*_{q}(\omega)).
\end{equation}
In this new basis, the inverse retarded Green's function is given by
\begin{align}
&\b{\tilde{P}}^R_{\mathrm{at}}(\omega)=(\b{\tilde{P}}^A_{\rm at}(\omega)^{-1})^\dagger\nonumber \\
&=\frac{1}{4}\bmx
4\delta\Sigma_\mathrm{at}-K_+(\omega) & -i(K_-(\omega)+2\omega) \\
i(K_-(\omega)+2\omega)  & -K_+(\omega)-2\omega_z.
\emx,
\end{align}
where $K_+(\omega)=K^A(-\omega)+K^R(\omega)$ and $K_-(\omega)=K^A(-\omega)-K^R(\omega)$. The inverse Keldysh Green's function is given by
\begin{align}
{\b{ \tilde{P}}}^K_{\mathrm{at}}(\omega)=
\frac{1}{4}\bmx
P^K_{+}(\omega)+4g(\omega) &   -iP^K_{-}(\omega) \\
 iP^K_{-}(\omega) & P^K_{+}(\omega)
\emx.
\label{DKnewbasis}
\end{align}
where $P^K_{+}(\omega)=P^K_\mathrm{at}(\omega)+P^K_\mathrm{at}(-\omega)$ and $P^K_-(\omega)=P^K_\mathrm{at}(\omega)-P^K_\mathrm{at}(-\omega)$. Again, we have simply made a change of basis without making any approximations [given Eq.~\eqref{eq:S_eff_b}]. 

Next we observe that the $\zeta$ fields are always gapped. Therefore, near the critical point, the low-frequency theory is completely governed by the $\phi$ fields. Integrating out the massive fields, we find the effective action for the $\zeta$ fields as
\begin{align}
S_{\phi}^{\rm eff}=
\int_\omega(\phi_{cl}^*,\phi^*_{q})
\bmx
0 &  P^A_{\phi}(\omega) \\
P^R_{\phi}(\omega) & P_{\phi}^{K}(\omega)
\emx
\left(\begin{array}{cc}
\phi_{cl} \\
\phi_{q}
\end{array}\right),
\label{c_baction}
\end{align}
where
\begin{align}
P^R_{\phi}(\omega)&=(P^A_{\phi}(\omega))^*\nonumber \\
=&\frac{1}{4}\bigg(4\delta\Sigma^R_\mathrm{at}-K^*_+(\omega)+\frac{(K_-(\omega)+2\omega)^2}{(K_+(\omega)-2\omega_z)}\bigg),
\label{GRCB}
\end{align}
and
\begin{align}
P_{\phi}^{K}(\omega)=\frac{1}{4}\bigg(4g(\omega)+P^K_{+}(\omega)\frac{|K_-(\omega)-2\omega|^2}{|K_+(\omega)-2\omega_b|^2}\nonumber \\
-2P^K_{+}(\omega)\mathrm{Re}\bigg[\frac{K_-^*(\omega)+2\omega}{K_+(\omega)-2\omega_z}\bigg]\bigg).
\label{GKCB}
\end{align}
The last term in Eq.~\eqref{GRCB} and the last two terms in Eq.~\eqref{GKCB} arise by integrating out the $\zeta$ fields. We now explicitly derive the low-frequency theory in two different cases and show that Green's functions find the same low-frequency description as the cavity mode near the critical point.

\subsection{Markovian bath on $\&$ non-Markovian bath on}
Here, we consider the case when both baths are present and $T_b=0$, $\mu_b\leq0$ (or, alternatively, $T_b\ne 0$ and $\mu_b< 0$). We then find the inverse retarded Green's function at low frequencies as
\begin{align}
P^R_{\phi}(\omega)\approx 
-r_{\mathrm{at}}+\bigg|\frac{\omega}{\omega_z}\bigg|^s(iv_{\mathrm{at},I}\sgn(\omega)-v_{\mathrm{at},R}),
\end{align}
where we have identified $r_{\mathrm{at}}\approx \delta y\frac{y_c\Delta}{\kappa^2+\Delta^2}$ near the critical point and we have defined
\begin{equation}
v_{\mathrm{at},R}=\frac{\pi \gamma}{4} (\csc(\pi s)+\cot(\pi s)),\quad v_{\mathrm{at},I}=\frac{\pi\gamma}{4}.
\end{equation}
The inverse Keyldsh Green's function in the low-frequency limit becomes 
\begin{align}
P_{\phi}^{K}(\omega)\approx g(\omega)\approx y_c^2\frac{i\kappa}{\Delta^2+\kappa^2}.
\label{equationphiK}
\end{align}
These are the same as the inverse retarded and Keldysh Green's functions presented in the main text up to multiplicative factors for the cavity mode [Eqs~\eqref{grcase3} and ~\eqref{DKcase1}]. We thus conclude that the low-frequency theory and the critical behavior is identical to the cavity mode.

\subsection{Markovian bath on $\&$ non-Markovian bath off}\label{sec:LEATOMICFIELDcaseIII}
In this limit (with $T_b=0$ and $\mu_b\leq0$), the low-frequency inverse retarded Green's function is
\begin{align}
P^R_{\phi}(\omega)\approx\delta\Sigma^R_\mathrm{at}\approx
-r_\mathrm{at}+\frac{iy_c^2\Delta \kappa}{(\Delta^2+\kappa^2)^2}\omega.
\end{align}
The inverse Keldysh Green's function is the same as the previous case [Eq.~\eqref{equationphiK}]. 
The above inverse retarded Green's function is similar to the inverse retarded Green's function of the cavity mode [Eq.~\eqref{GREENSFUNCTIONMBONNMBOFF}]. 
Also, similar to the cavity mode [Eq.~\eqref{GREENSFUNCTIONMBONNMBOFF}], the inverse Keyldsh Green's function of the atomic field is a constant. In fact, the effective low-frequency theory of the atomic field turns out to be the same as that of the cavity mode (up to multiplicative factors); see Sec.~\ref{CASEMBONNMBOFF}. Specifically, the effective temperature of the atomic field (obtained from the distribution function in the low-frequency limit) coincides with that of the photonic mode, $T_{\rm eff}=\frac{\kappa^2+\Delta^2}{4\Delta}$; cf. Sec.~\ref{sec:othermodelscaseIII}.

\section{Interactions}
In this section, we present the details of integrating out either the cavity field or atomic field at finite $N$.

\subsection{Integrating out the photonic field in the presence of interactions}\label{sec:INTERACTIONS}
In this section, we present the details of integrating out the cavity field at finite $N$. The interaction term between the cavity and atomic fields [see Eqs.~\eqref{sab} and~\eqref{sint}] can be conveniently rewritten as
\begin{align}
&S_{\rm ph-at}+S_{\rm int}=\nonumber \\
&\int_t(a_q(t)+a^*_q(t))h(t)+
(a_{cl}(t)+
a^*_{cl}(t))f(t),
\label{INTABINT}
\end{align}
with 
\begin{align}
f(t)&=-\frac{y}{2}(b^*_{q}(t)+b_{q}(t))\nonumber \\
&+\frac{y}{8N}\bigg[(b^*_{q}(t)+b_{q}(t))(b^*_{cl}(t)b_{cl}(t)+b^*_{q}(t)b_{q}(t))\nonumber \\
&~~~~+(b^*_{cl}(t)+b_{cl}(t))(b^*_{cl}(t)b_{q}(t)+b^*_{q}(t)b_{cl}(t))\bigg],
\end{align}
being real and $h(t)$ obtained by swapping classical and quantum fields in $f(t)$. We can exactly integrate out the cavity mode to arrive at the effective action for the $b$ field as the action is quadratic in $a$ (this is true to all orders of $1/N$; however, we are only interested in the lowest-order terms). 
We find (see Ref.~\cite{2016RPPh...79i6001S})
\begin{align}
&S_\mathrm{at}^{\rm eff}=S_\mathrm{at,0}^{\rm eff}-\nonumber \\
\!\!&\int_{t,t'} (f(t),h(t))\bmx
G^K_{\mathrm{ph}}(t-t') & G^R_{\mathrm{ph}}(t-t')   \\
G^A_{\mathrm{ph}}(t-t') & 0
\emx\vect{f(t')}{h(t')},
\end{align}
where $S_\mathrm{at,0}^{\rm eff}$ is given by Eq.~\eqref{eq:S_eff_b} with $y=0$ (without absorbing a factor of $1/2$). Since we are only concerned with the slow part of $h(t)$ and $f(t)$ compared to the (bare) Green's functions, we have
\begin{align}
S_\mathrm{at}^{\rm eff}\approx &\, S_\mathrm{at,0}^{\rm eff}-\int_tf^2(t)\int_{t'}G^K_{\mathrm{ph}}(t-t') \nonumber\\
&\,-\int_t f(t)h(t)\int_{t'}\left[G^R_{\mathrm{ph}}(t-t')+G^A_{\mathrm{ph}}(t-t')\right]\nonumber \\
=&\,S_\mathrm{at,0}^{\rm eff}+\frac{2}{\Delta^2+\kappa^2}\int_t\left[i\kappa f^2(t)+\Delta f(t)h(t)\right].
\label{SINTLOWTIME}
\end{align}
To proceed, we express Eq.~\eqref{SINTLOWTIME} in terms of the real-valued $\phi$ and $\zeta$ fields. Using our basis transformation of the atomic fields, $b_{cl/q}(t)=\frac{1}{2}(\phi_{q}(t)-i\zeta_{q}(t))$, 
introduced in Appendix~\ref{sec:LEATOMICFIELD} [see Eq.~\eqref{bfieldtransform1}], we find that $f(t)$ is given by
\begin{align}
f(t)=-\frac{y}{2}\phi_{q}(t)+\frac{y}{16N}\bigg(\frac{1}{2}\phi_{q}(t)\big(\phi_{cl}^2(t)+\zeta_{cl}^2(t)+\nonumber \\
\phi_{q}^2(t)+ \zeta_{q}^2(t)\big)+(\phi_{cl}(t)\zeta_{cl}(t)\zeta_{q}(t)+\phi^2_{cl}(t)\phi_{q}(t))\bigg),
\label{ftexpression}
\end{align}
and $h(t)$ is obtained by switching the classical and quantum field labels in $f(t)$. Integrating out the $\zeta$ field (which simply renormalizes the Green's functions by terms of the order $1/N$ that can be neglected), keeping the most relevant terms in the renormalization-group sense and finally taking the low-frequency limit\footnote{Here, we have explicitly assumed that the classical field has a larger scaling dimension than the quantum field. This is not the case when $\kappa=0$, i.e., in the absence of the Markovian bath. However, our finite-size scaling analysis is valid in the $\kappa=0$ case too since the interaction terms are proportional to $1/N$.} gives $S_\mathrm{phit}^{\rm tot} \approx S^{\rm eff}_\phi+S^{\rm eff}_{\rm \phi, int}$, where $S^{\rm eff}_{\phi}$ is given by Eq.~\eqref{c_baction}, and
\begin{align}
S^{\rm eff}_{\rm \phi, int} =-\frac{y_c^2}{2N}\frac{\Delta}{\Delta^2+\kappa^2}\int_t \phi_{q}(t)\phi_{cl}^{3}(t). 
\label{gatomicN}
\end{align}
This equation determines $g_\mathrm{at}=\frac{y_c^2}{2}\frac{\Delta}{\Delta^2+\kappa^2}$ and is the main result of this part of the Appendix. Note that we have absorbed factors of $1/2$ into the $\phi$ fields in the interaction term to match the notation of the previous section.

\subsection{Integrating out the atomic field in the presence of interactions}\label{sec:INTERACTIONSAEFF}
In this section, we present the details of integrating out the atomic field at finite $N$. It is helpful to express the action in terms of $x$, $z$, $\phi$ and $\zeta$ fields as there are several simplifications that occur. We first note that the interaction given in Eq.~\eqref{INTABINT} does not depend on the $z$ fields, thus they can be easily integrated out. In doing so, we obtain an effective action involving three fields, $S^{\rm eff} \approx S^{\rm eff}_{x,0}+S_\mathrm{at}+S_{\rm int}+S_{\rm ph-at}^{\phi},$ where $S^{\rm eff}_{x,0}$ is given by Eq.~\eqref{c_aaction} with $y$ set to zero\footnote{\label{footnote7}Here, we have not yet absorbed the 1/2 into the fields as done in Sec.~\ref{sec:LEFT} of the main text and Sec.~\ref{sec:LEATOMICFIELD} of the Appendix.} and $S_{\rm ph-at}^{\phi}$ is given by Eq.~\eqref{sab} with the original fields written in terms of the rotated fields. Further simplification is possible because the $\zeta$ field does not enter in the coupling between the $a$ and $b$ fields and the coupling between the $\phi$ and $\zeta$ vanishes with the frequency [see Eq.~\eqref{eq:S_eff_b_newbasis}]. A consequence of the latter is that the terms generated by integrating out the $\zeta$ field  are less relevant in the renormalization-group sense than the terms in $S_{\rm int}$ that do not depend on $\zeta$ [see, for example, Eq.~\eqref{ftexpression}]. This means that we can integrate out the $z$ field and neglect the terms in $S_{\rm int}$ that depend on $\zeta$. Doing so gives an effective action that depends only on $x$ and $\phi$ fields, $S^{\rm eff} \approx S^{\rm eff}_{x,0}+S^{\rm eff}_\mathrm{\phi,0}+S^{\phi}_{\rm int}+S^\phi_{\rm ph-at}$. Here, $S^{\rm eff}_{\phi,0}$ is given by Eq.~\eqref{c_baction}  with $y$ set to zero\textsuperscript{\ref{footnote7}} with $S^{\phi}_{\rm int}$
indicating those terms in $S_{\rm int}$ that only include the $\phi$ fields.

We now integrate out the $\phi$ field in order to obtain an effective action for the $x$ field. Expanding the partition function to order $1/N$ (higher-order terms in $1/N$ are discussed at the end of this section) gives
\begin{align}
Z\approx
&\int d [x,\phi]e^{iS^{\rm eff}_{x,0}+iS^{\rm eff}_{\phi,0}+\frac{i}{2}\int_t \phi_q(t)J_{cl}(t)+\phi_{cl}(t)J_{q}(t)}\nonumber \\
\times\bigg(&1+\frac{iy}{32N}\int_t \big[x_{q}(t)(3\phi_{cl}(t)\phi^2_{q}(t)+\phi_{cl}^3(t))+\nonumber \\
&x_{cl}(t)(3\phi_q(t)\phi^2_{cl}(t)+\phi_q^3(t))\big]\bigg).
\end{align}
Here, $S^{\phi}_{\rm int}$ has been explicitly written out and $J_{cl/q}(t)=-yx_{cl/q}(t)$. Integrating out $\phi$, this can be written as
\begin{align}
&Z\approx\int d [x]e^{iS^{\rm eff}_{x,0}}\bigg(Z_{\phi}+\frac{y}{32N}\int_t (x_q(t)(3\frac{\delta^3 Z_{\phi}}{\delta J_{q}(t)\delta^2 J_{cl}(t)}\nonumber \\
&+\frac{\delta^3 Z_{\phi}}{\delta^3 J_{q}(t)})+
x_{cl}(t)(3\frac{\delta^3 Z_{\phi}}{\delta J_{cl}(t)\delta^2 J_{q}(t)}+\frac{\delta^3 Z_{\phi}}{\delta^3 J_{cl}(t)}))\bigg),
\label{totalpartition}
\end{align}
where
\begin{align}
Z_{\phi}&=\int d[\phi]e^{iS^{\rm eff}_{\phi,0}+\frac{i}{2}\int_t \phi_{q}(t)J_{cl}(t)+\phi_{cl}(t)J_{q}(t)} \nonumber\\ 
&=\exp\bigg\{-i\int_t\int_{t'}\big[ J_{q}(t)G^{K}_{\rm at}(t-t')J_{q}(t')\nonumber \\ 
&\qquad +2J_{q}(t)G^{R}_{\rm at}(t-t')J_{cl}(t')\big]\bigg\}.
\end{align}
Here, we have used the fact that $4G^{R,A,K}_{\rm at}=G^{R,A,K}_\phi$ when $y=0$ and $G_{\rm at}^R(t)=G_{\rm at}^A(-t)$. Keeping only the most relevant terms in the derivatives of $Z_{\phi}$ (also neglecting $1/N$ corrections in the retarded and advanced Green's functions) and absorbing factors of 1/2 (see the beginning of Sec.~\ref{sec:LEFT}) yields
\begin{align}
Z\approx\int d [x]e^{iS^{\rm eff}_{x}}\bigg(1-\frac{iy^4}{2N\omega_z^3}\int_t x_q(t)x_{cl}^3(t)\bigg),
\label{PARTION_X}
\end{align}
where $S^{\rm eff}_x$ is given by Eq.~\eqref{c_aaction} in the main text and we have used the fact that the retarded and advanced Green's function of the atomic field at zero frequency are $-1/\omega_z$. We have also used the fact that only the second and third terms (under the time integral) in Eq.~\eqref{totalpartition} generate terms with three classical fields. Replacing $y$ by $y_c$ near the critical point, we can exponentiate the expression in \cref{PARTION_X} to obtain an effective interaction term in the action as
\begin{equation}
S^{\rm eff}_{x,\rm int}=-\frac{(\Delta^2+\kappa^2)^2}{2N\Delta^2\omega_z}\int_t x_q(t)x_{cl}^3(t). 
    \label{seffxappend}
\end{equation}
From this equation, we can then identify $g_{\rm ph}=\frac{(\Delta^2+\kappa^2)^2}{2\Delta^2\omega_z}$.

The exponentiation described above requires a careful treatment of the higher-order terms in $1/N$ in the partition function. In general, one must compute
\begin{align*}
    &\int [d\phi]e^{i (S^{\rm eff}_{\phi,0}+S^\phi_{\rm ph-at}+S^\phi_{\rm int})}=
    \nonumber \\
     &\int [d\phi]e^{i(S^{\rm eff}_{\phi,0}+S^\phi_{\rm ph-at})}\sum_{n=0}^\infty \frac{i^n}{n!}(S^\phi_{\rm int})^n\equiv \sum_{n=0}^\infty \frac{i^n}{n!}\langle(S^\phi_{\rm int})^n\rangle_\phi.
 \end{align*}
But we shall argue that the most dominant contribution to each term in the expansion comes from the disconnected correlations, that is, $\langle(S^\phi_{\rm int})^n\rangle_\phi\approx\big(\langle S^\phi_{\rm int}\rangle_\phi\big)^n$. To see this, let us 
consider $n=2$, for example, and write the action as $S^\phi_{\rm int}=\frac{1}{N}\int_t L(t)$ with $L(t)$ the (Keldysh) Lagrangian; we have also factored out the coefficient $1/N$. For $n=2$, we have 
\begin{align*}
\langle (S^\phi_{\rm int})^2\rangle_\phi 
&=(\langle S^\phi_{\rm int}\rangle_\phi)^2+ \langle\!\langle (S^\phi_{\rm int})^2\rangle\!\rangle_\phi \\
&\approx\frac{1}{N^2}\int_{t,t'} L_1(t)L_1(t')+ \frac{1}{N^2}\int_t L_2(t),    
\end{align*}
for some local functions $L_1(t)$ and $L_2(t)$ that only depend on the $x$ fields; we have used $\langle\!\langle \cdot\rangle\!\rangle$ to denote the connected correlation. 
Notice that the second term in the last line of this equation is local in time due to the short-range correlation of the Gaussian $\phi$ correlators. This should be contrasted with the first term that is highly non-local in time. This observation is the reason why the disconnected term is more relevant. A similar argument can be extended to any $n$. The partition function can be then conveniently exponentiated as
\begin{align*}
    Z=\int d [x]e^{iS^{\rm eff}_{x}}\sum_{n=0}^\infty \frac{i^n}{n!}(\langle S^\phi_{\rm int}\rangle_\phi)^2=\int d [x]e^{iS^{\rm eff}_{x}+i\langle S^\phi_{\rm int}\rangle_\phi}.
\end{align*}
This completes our derivation of the effective interaction term in the action in \cref{seffxappend}.

\subsection{Mean-field equation}\label{sec:INTERACTIONSMEANFIELD}
In this section, we compare the interaction coefficients, $g_{\rm ph}$ and $g_{\rm at}$, obtained by integrating out the fields to the ones obtained by deriving the mean-field equations. As a nontrivial check, we show that they indeed agree. We begin by taking the derivative of the Keldysh-Schwinger action [see \cref{FULLACT,sint}] with respect to the quantum fields in the original basis,
\begin{align}
    \frac{\partial S}{\partial a^*_{q}(t)}=\int_{t'}[P^R_{\rm ph}(t,t')a_{cl}(t')+
    P^K_{\rm ph}(t,t')a_{q}(t')]+g(t)=0,
\end{align}
and
\begin{align}
    \frac{\partial S}{\partial b^*_{q}(t)}=\int_{t'}[P^R_{\rm at}(t,t')b_{cl}(t')+P^K_{\rm at}(t,t')b_{q}(t')]+\nonumber \\ 2\mathrm{Re}[a_{cl}(t)] \frac{\partial f(t)}{\partial b^*_{q}(t)}+ 2\mathrm{Re}[a_{q}(t)] \frac{\partial g(t)}{\partial b^*_{q}(t)}=0.
\end{align}
Setting the expectation value of the quantum fields to zero and assuming a constant value for the classical fields, we find
\begin{align}
    P^R_{\rm ph}(\omega=0)a_{cl}-y\mathrm{Re}[b_{cl}](1-\frac{|b_{cl}|^2}{4N})=0,
\end{align}
and
\begin{align}
    P^R_{\rm at}(\omega=0)b_{cl}-y\mathrm{Re}[a_{cl}]\big(1-
    \frac{1}{2N}(b_{cl}\mathrm{Re}[b_{cl}]+\frac{|b_{cl}|^2}{2})\big)=0.
\end{align}
Rewriting these equations in the rotated basis and assuming that the atomic field is real  \cite{PhysRevA.87.063622}, we find
\begin{align}
    \frac{P^R_{\rm ph}(\omega=0)}{2i\sin p}(x_{cl}e^{ip}-z_{cl})-\frac{y}{2}\phi_{cl}+
    \frac{y}{32N}\phi_{cl}^3=0,
\end{align}
and
\begin{align}
    \frac{P^R_{\rm at}(\omega=0)}{2}\phi_{cl}+x_{cl}\big(-\frac{y}{2}+  \frac{3y}{32N}\phi^2_{cl}\big)=0.
\end{align}
Solving for $\phi_{cl}$ (in the large $N$ limit and with $z_{cl}=0$), we obtain 
\begin{equation}
\big(-\frac{\omega_z}{2}+\frac{1}{2}\frac{\Delta y^2}{\Delta^2+\kappa^2}\big)\phi_{cl}-\frac{y^2}{8N}\frac{\Delta}{\Delta^2+\kappa^2}\phi_{cl}^3=0.
 \end{equation}
Near the critical point (and absorbing factors of 1/2 as done in the previous section), we have $r_{\rm at}\phi_{cl}+\frac{g_{\rm at}}{N}\phi_{cl}^3=0$. Thus, the interaction coefficient obtained in the mean-field equation is in agreement with the one obtained by formally integrating out the photonic field (Eq.~\eqref{gatomicN}). Solving for $x_{cl}$ gives
 \begin{equation}
     \big(-1+\frac{2\Delta y^2}{(\Delta^2+\kappa^2)\omega_z}\big)x_{cl}-\frac{y^4\Delta}{4\omega^3_z(\Delta^2+\kappa^2)}x^3_{cl}=0.
 \end{equation}
 Near the critical point, we have $r x_{cl}+\frac{g_{\rm ph}}{N}x_{cl}^3=0$. Thus, the interaction coefficient for the $x$ field obtained from the mean-field equation is in agreement with the one obtained by formally integrating out the atomic field, Eq.~\eqref{seffxappend}, as expected.
 \section{Fokker-Planck equation}\label{sec:FPEQ}
 In this section, we investigate the Fokker-Planck (FP) equation that we have conjectured in the main text [Eq.~\eqref{FPEQ}],
 \begin{align}
     \partial_tP(x,t)=
     Ar\partial_t^{1-s}\partial_x(xP)+B\partial_t^{2(1-s)}(\partial_x^2P),
 \end{align}
for $s>1/2$; here, $P=P(x,t)$ is the probability distribution at time $t$. We assume that the time starts at $t=0$, as such our fractional derivatives are re-defined as 
 \begin{align}
\begin{split}
\partial_t^sf(t)&=\frac{1}{\Gamma(1-s)}\frac{d}{dt}\int_{0}^t\frac{f(t')}{(t-t')^s}dt'.
\end{split}
\end{align}
We will see that this equation reproduces the scaling of various correlation functions discussed in the main text.
 
We first consider the critical point, $r=0$. In this limit, we simply recover the fractional FP equation for a free particle (i.e., not bound to a potential) that exhibits anomalous diffusion \cite{Metzler99}, which directly reproduces the correct square-mean-displacement. For completeness, we explicitly show how to solve the FP equation in this limit. Using various fractional-calculus identities \cite{kilbas2006theory} and Fourier transforming (in position space), we find
 \begin{align}
     \partial^{2s-1}_t P(k,t)= -Bk^2P(k,t).
 \end{align}
 The exact solution to this equation is given by \cite{kilbas2006theory}
 \begin{equation}
     P(k,t)=bt^{2s-1}E_{2s-1,2s-1}(Bk^2 t^{2s-1}),
 \end{equation}
 where $E$ is the generalized Mittag-Leffler function and $b=\partial_t^{2s-2}(P(k,0))$. $b$ serves as the initial condidtion and is conveniently assumed to be independent of $k$. Fourier transforming $P(k,t)$ gives
  \begin{align}
     P(x,t)=\int_{-\infty}^{\infty} \frac{dk}{2\pi} P(k,t)e^{ikx}=\nonumber \\
     \frac{1}{2\pi}\frac{b}{ x}H^{1,0}_{1,1}\bigg[\frac{x^2}{Bt^{2s-1}}\bigg|^{(2s-1,2s-1)}_{(1,2)}\bigg],
 \end{align}
 where $H$ is the Fox H-function. For $s=1$, this reduces to Gaussian diffusion as expected. For $s\neq1$, even taking the long-time limit of this function proves to be difficult \cite{kilbas2006theory}, however, a simple rescaling yields 
 \begin{equation}
     \langle x^2(t)\rangle=\int dx x^2 P(x,t)\propto  t^{2s-1}.
 \end{equation}
 This correctly reproduces Eq.~\eqref{Eq. anamolous} in the main text.
 
 We now turn to finite $r$. Unfortunately, we are not able to directly solve Eq.~\eqref{FPEQ} as done for $r=0$. Instead, we resort to a scaling argument to calculate critical exponents. We first note that there is only one time scale and one ``length'' scale (corresponding to $x$) imposed by Eq.~\eqref{FPEQ}. This means $P(x,t) \sim {\cal P}(r^{1-\frac{1}{2s}}x,r^{\frac{1}{s}} t)$ up to a normalization coefficient (we do not keep track of the coefficients $A$ and $B$ for convenience). Here, we have assumed sufficiently long times such that the system retains no memory of initial  conditions. We should however normalize the probability distribution, $\int dx P(x,t)=1$, which then yields the normalized probability distribution function
 \begin{align}
P(x,t)=r^{1-\frac{1}{2s}} {\cal N}(r^{\frac{1}{s}}t){\cal P}(r^{1-\frac{1}{2s}}x,r^{\frac{1}{s}}t),
\end{align}
where ${\cal N}(t')= \int dx' {\cal P}(x',t')$.

Next, we characterize the fluctuations $\langle x^2\rangle$ in the steady state. In this limit ($t\rightarrow\infty$), we expect $P(x,t)$ to become independent of $t$ and approach a stationary function, $P_{\rm st}(x)$. We then obtain
\begin{align}
P_{\rm st}(x)=\lim_{t\rightarrow\infty}P(x,t)= 
r^{1-\frac{1}{2s}} {\cal P}_{\mathrm{st}}(r^{1-\frac{1}{2s}}x),
\end{align}
 where  ${\cal P}_{\mathrm{st}}(x')=\lim_{t'\rightarrow\infty} {\cal N}(t') {\cal P}(x',t')$.
With the above scaling function, we can then obtain the scaling of the fluctuations in the steady state by simply rescaling $x$:
 \begin{equation}
     \lim_{t\rightarrow\infty}\langle x^2(t)\rangle=\int dx x^2 P_{\rm st}(x)\propto \frac{1}{r^{2-\frac{1}{s}}}.
 \end{equation}
Indeed, the exponent is identical to \cref{correlationEXPmaincase} in the main text; see also Table~\ref{CritEXPTABLE}. This gives further credence to our conjectured Fokker-Planck equation.

\twocolumngrid
\bibliography{non_markovian.bib}

\end{document}